\documentclass[fleqn,usenatbib]{mnras}

\usepackage{newtxtext,newtxmath}

\usepackage[T1]{fontenc}
\usepackage{ae,aecompl}
\usepackage{graphicx}	
\usepackage{amsmath}	
\usepackage{amssymb}	
\usepackage{xcolor}
\usepackage{gensymb}
\usepackage{chngcntr}

\usepackage{subcaption}  
\usepackage{url}

\usepackage [autostyle, english = british]{csquotes}  

\title[Galaxy Classification with Active Learning]{{Galaxy Zoo: Probabilistic Morphology through Bayesian CNNs and Active Learning}}

\author[M. Walmsley et al.]{
Mike Walmsley$^{1}$\thanks{E-mail: mike.walmsley@physics.ox.ac.uk (MW)},
Lewis Smith$^{2}$,
Chris Lintott$^{1}$, 
Yarin Gal$^{2}$, 
Steven Bamford$^{3}$, 
\newauthor
Hugh Dickinson$^{4,8}$, 
Lucy Fortson$^{4,8}$, 
Sandor Kruk$^{5}$, 
Karen Masters$^{6,7}$,
Claudia Scarlata$^{4,8}$, 
\newauthor
Brooke Simmons$^{9,10}$,
Rebecca Smethurst$^{1}$, 
Darryl Wright $^{4,8}$
\\  
$^{1}$Oxford Astrophysics, Department of Physics, University of Oxford, Denys Wilkinson Building, Keble Road, Oxford, OX1 3RH, UK\\
$^{2}$Oxford Computer Science, University of Oxford, 15 Parks Rd, Oxford, OX1 3QD, UK \\
$^{3}$School of Physics and Astronomy, University of Nottingham, University Park, Nottingham NG7 2RD, UK\\
$^{4}$School of Physics and Astronomy, University of Minnesota, 116 Church St SE, Minneapolis, MN 55455, USA\\
$^{5}$European Space Agency, ESTEC, Keplerlaan 1, NL-2201 AZ, Noordwijk, The Netherlands\\
$^{6}$Haverford College, Department of Physics and Astronomy, 370 Lancaster Avenue, Haverford, Pennsylvania 19041, USA\\
$^{7}$Institute for Cosmology and Gravitation, University of Portsmouth, Dennis Sciama Building, Burnaby Road, Portsmouth, PO1 3FX, UK\\
$^{8}$Minnesota Institute for Astrophysics, University of Minnesota, Minneapolis, MN 55455, USA\\
$^{9}$Physics Department, Lancaster University, Lancaster, LA1 4YB, UK\\
$^{10}$Center for Astrophysics and Space Sciences (CASS), Department of Physics, University of California, San Diego, CA 92093, USA\\
}

\date{Accepted XXX. Received YYY; in original form ZZZ}

\pubyear{2019}

\begin{document}
\label{firstpage}
\pagerange{\pageref{firstpage}--\pageref{lastpage}}
\maketitle

\begin{abstract}
    We use Bayesian convolutional neural networks and a novel generative model of Galaxy Zoo volunteer responses to infer posteriors for the visual morphology of galaxies.
    Bayesian CNN can learn from galaxy images with uncertain labels and then, for previously unlabelled galaxies, predict the probability of each possible label.
    Our posteriors are well-calibrated (e.g. for predicting bars, we achieve coverage errors of 11.8\% within a vote fraction deviation of 0.2) and hence are reliable for practical use.
    Further, using our posteriors, we apply the active learning strategy BALD to request volunteer responses for the subset of galaxies which, if labelled, would be most informative for training our network.
    We show that training our Bayesian CNNs using active learning requires up to 35-60\% fewer labelled galaxies, depending on the morphological feature being classified.
    By combining human and machine intelligence, Galaxy Zoo will be able to classify surveys of any conceivable scale on a timescale of weeks, providing massive and detailed morphology catalogues to support research into galaxy evolution.
\end{abstract}




\section{Introduction}
\label{introduction}

Galaxy Zoo was created because SDSS-scale surveys could not be visually classified by professional astronomers \citep{Lintott2008}.
In turn, Galaxy Zoo is being gradually outpaced by the increasing scale of modern surveys like DES \citep{Flaugher2005}, PanSTARRS \citep{Kaiser2010}, the Kilo-Degree Survey \citep{DeJong2015}, and Hyper Suprime-Cam \citep{Aihara2018}.

Each of these surveys can each image galaxies as fast or faster than those galaxies are being classified by volunteers.
For example, DECaLS \citep{Dey2018} contains (as of Data Release 5) approximately 350,000 galaxies suitable for detailed morphological classification (applying $r$ < 17 and \texttt{petroR90\_r}\footnote{\texttt{petroR90\_r} is the Petrosian radius which contains 90\% of the $r$-band flux} > 3 arcsec, the cuts used for Galaxy Zoo 2 in \citealt{Willett2013}).
Collecting 40 independent volunteer classifications for each galaxy, as for Galaxy Zoo 2 \citep{Willett2013}, would take approximately five years at the current classification rate.
The Galaxy Zoo science team must therefore both judiciously select which surveys to classify and, for the selected surveys, reduce the number of independent classifications per galaxy.
The speed at which we can accurately classify galaxies severely limits the scale, detail, and quality of our morphology catalogues, diminishing the scientific value of such surveys.

The next generation of surveys will make this speed limitation even more stark.
Euclid\footnote{15,000 deg$^2$ at 0.30$^{\prime\prime}$ half-light radius PSF from 2022, \citealt{Laureijs2011}}, LSST\footnote{18,000 deg$^2$ to 0.39$^{\prime\prime}$ half-light radius PSF from 2023, \citealt{2009arXiv0912.0201L}} and WFIRST \footnote{2,000 deg$^2$ at 0.12$^{\prime\prime}$ half-light radius PSF from approx. 2025, \citealt{Spergel2013}} are expected to resolve the morphology of unprecedented numbers of galaxies.
This could be revolutionary for our understanding of galaxy evolution, but only if such galaxies can be classified.
The future of morphology research therefore inevitably relies on automated classification methods.
Supervised approaches (given human-labelled galaxies, predict labels for new galaxies) using convolutional neural networks (CNNs) are increasingly common and effective \citep{Cheng2019}.
CNNs outperform previous non-parametric approaches \citep{Dieleman2015, Huertas-Company2015a}, and can be rapidly adapted to new surveys \citep{Sanchez2018} and to related tasks such as light profile fitting \citep{Tuccillo2017}.
Unsupervised approaches (cluster examples without any human labels) also show promise \citep{Hocking2015}.\\

However, despite major progress in raw performance, the increasing complexity of classification methods poses a problem for scientific inquiry.
In particular, CNNs are `black box' algorithms which are difficult to introspect and do not typically provide estimates of uncertainty.
In this work, we combine a novel generative model of volunteer responses with Monte Carlo dropout \citep{Gal2017} to create Bayesian CNNs that predict \textit{posteriors} for the morphology of each galaxy.
Posteriors are crucial for drawing statistical conclusions that account for uncertainty, and so including posteriors significantly increases the scientific value of morphology catalogues.
Our Bayesian CNNs can predict posteriors for surveys of any conceivable scale.

Limited volunteer classification speed remains a hurdle; we need to collect enough responses to train our Bayesian networks. 
How do we train Bayesian networks to perform well while minimising the number of new responses required?
Recent work suggests that transfer learning \citep{Bello2017} may be effective.
In transfer learning, models are first trained to solve similar tasks where training data is plentiful and then `fine-tuned' with new data to solve the task at hand.
Results using transfer learning to classify new surveys, or to answer new morphological questions, suggest that models can be fine-tuned using only thousands \citep{Ackermann2018, Khan2018} or even hundreds \citep{DominguezSanchez2019} of newly-labelled galaxies, with only moderate performance losses compared to the original task. 

Each of these authors randomly selects which new galaxies to label.
However, this may not be optimal.
Each galaxy, if labelled, provides information to our model; they are \textit{informative}.
Our hypothesis is that all galaxies are informative, but some galaxies are more informative than others.
We use our galaxy morphology posteriors to apply an active learning strategy \citep{Houlsby2011}: \textit{intelligently selecting the most informative galaxies for labelling by volunteers}.
By prioritizing the galaxies that our strategy suggests would, if labelled, be most informative to the model, we can create or fine-tune models with even fewer newly-labelled data.

In the first half of this work (Section \ref{methods}), we present Bayesian CNNs that predict posteriors for the morphology of each galaxy.
In the second (Section \ref{methods_active_learning}), we simulate using our posteriors to select the most informative galaxies for labelling by volunteers.

\section{Posteriors for Galaxy Morphology}
\label{methods}

A vast number of automated methods have been used as proxies for `traditional' visual morphological classification. 
Non-parametric methods such as CAS \citep{Conselice2003} and Gini \citep{Lotz2004} have been commonly used, both directly and to provide features which can be used by increasingly sophisticated machine learning strategies \citep{Scarlata2006, Banerji2010, Huertas-Company2010, Freeman2013, Peth2016}. 
Most of these methods provide imperfect proxies for expert classification \citep{Lintott2008}.
The key advantage of CNNs is that they learn to approximate human classifications directly from data, without the need to hand-design functions aimed at identifying relevant features \citep{LeCun2015}.
CNNs work by applying a series of spatially-invariant transformations to represent the input image at increasing levels of abstraction, and then interpreting the final abstraction level as a prediction.
These transformations are initially random, and are `learned' by iteratively minimising the difference between predictions and known labels.
We refer the reader to \cite{LeCun2015} for a brief introduction to CNNs and to \cite{Dieleman2015}, \cite{Lanusse2017}, \cite{Kim2017} and \cite{Hezaveh2017} for astrophysical applications. 

Early work with CNNs immediately surpassed non-parametric methods in approximating human classifications \citep{Huertas-Company2015a, Dieleman2015}.
Recent work extends CNNs across different surveys \citep{Sanchez2018, Khan2018} or increasingly specific tasks \citep{Sanchez2017,Tuccillo2017,HuertasCompany2018,Walmsley2018}.
However, these previous CNNs do not account for uncertainty in training labels, limiting their ability to learn from all available data (one common approach is to train only on `clean' subsets). 
Previous CNNs are also not designed to make probabilistic predictions (though they have been interpreted as such), limiting the reliability of conclusions drawn using such methods (see Appendix A).

Here, we present Bayesian CNNs for morphology classification. 
Bayesian CNNs provide two key improvements over previous work:
\begin{enumerate}
    \item We account for varying (i.e. heteroskedastic) uncertainty in volunteer responses
    \item We predict full posteriors over the morphology of each galaxy
\end{enumerate}

We first introduce a novel framework for thinking about Galaxy Zoo classifications in probabilistic terms, where volunteer responses are drawn from a binomial distribution according to an unobserved (latent) parameter: the `typical' response probability (Section \ref{probabilistic_framework}).
We use this framework to construct CNNs that make probabilistic predictions of Galaxy Zoo classifications (Section \ref{probabilistic_prediction_with_cnn}).
These CNNs predict a typical response probability for each galaxy by maximising the likelihood of the observed responses.
By maximising the likelihood, they learn effectively from heteroskedastic labels; the likelihood reflects the fact that more volunteer responses are more indicative of the `typical` response than fewer responses.
To account for the uncertainty in the CNN weights, we use Monte Carlo dropout \citep{Gal2017} to marginalise over possible CNNs (Section \ref{monte_carlo_dropout}).
Our final predictions (Section \ref{probabilistic_results}) are posteriors of how a typical volunteer would have responded, had they been asked about each galaxy.
These can then be used to classify surveys of any conceivable scale (e.g. LSST, Euclid), helping researchers make reliable inferences about galaxy evolution using millions of labelled galaxy images.\\

\subsection{Probabilistic Framework for Galaxy Zoo}
\label{probabilistic_framework}

\begin{figure}
    \includegraphics[width=\columnwidth]{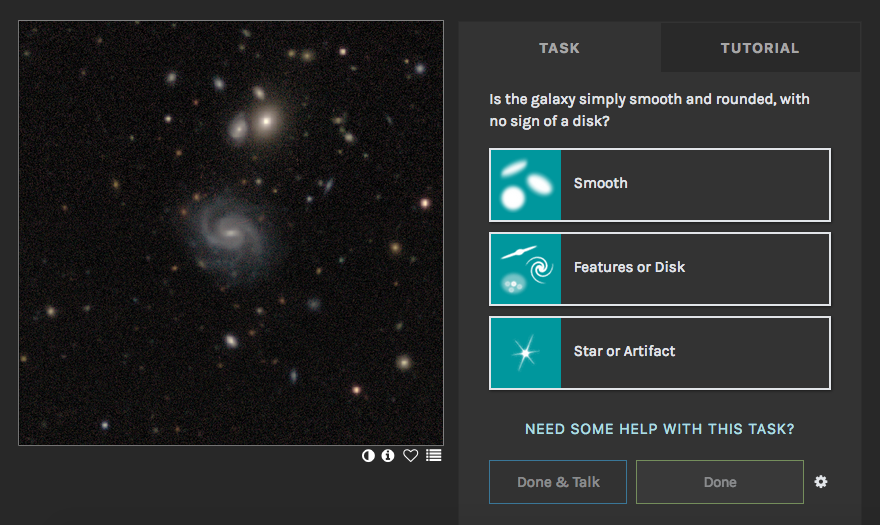}
    \caption{
        The Galaxy Zoo web interface as shown to volunteers.
        This screenshot shows the first question in the decision tree: is the galaxy smooth or featured?}
        \label{galaxy_zoo_web_interface}
\end{figure}

\href{www.galaxyzoo.org}{Galaxy Zoo} asks members of the public to volunteer as `citizen scientists' and label galaxy images by answering a series of questions.
Figure \ref{galaxy_zoo_web_interface} illustrates the web interface.

We aim to make a probabilistic prediction for the response of a typical volunteer.
To do this, we need to model how each volunteer response is generated.
Formally, each Galaxy Zoo decision tree question asks $N_i$ volunteers to view galaxy image $x_{i}$ and select the most appropriate answer $A_j$ from the available answers $\{A\}$.
This reduces to a binary choice; where there are more than two available answers ($|\{A\}| > 2$), we can consider each volunteer response as either $A_j$ (positive response) or not $A_j$ (negative response).
We can therefore apply our model to questions with any number of answers. 

Let $k_{ij}$ be the number of volunteers (out of $N_i$) observed to answer $A_j$ for image $x_{i}$.
We assume that there is a true fraction $\rho_{ij}$ of the population (i.e. all possible volunteers) who
would give the answer $A_j$ for image $x_i$. 
We assume that volunteers are drawn uniformly from this population, 
so that if we ask $N_i$ volunteers about image $x_i$, we expect that the distribution over the number of positive answers $k_{ij}$ to be binomial:

\begin{equation}
    k_{ij} \sim \text{Bin}(\rho_{ij}, N_i) 
\end{equation}
\begin{equation}
    \label{k_distribution}
    p(k_{ij}| x_{ij}, N_i) = {N_i \choose k_i} \rho_{ij}^{k_{ij}} (1 - \rho_{ij})^{N_i-k_{ij}}
\end{equation}

This will be our model for how each volunteer response $k_{ij}$ was generated.
Note that $\rho_{ij}$ is a latent variable: we only observe the responses $k_{ij}$, never $\rho_{ij}$ itself.

\subsection{Probabilistic Prediction with CNNs}
\label{probabilistic_prediction_with_cnn}

Having established a novel generative model for our data, we now aim to infer the likelihood of observing a particular $k$ for each galaxy $x$ (for brevity, we omit subscripts).

Let us consider the scalar output from our neural network $f^{w}(x)$ as a (deterministic) prediction for $\rho$, and hence a probabilistic prediction for $k$:

\begin{equation}
    \label{model_likelihood}
    p(k|x, w) = \text{Bin}(k|f^w(x), N)
\end{equation}

For each labelled galaxy, we have observed $k$ positive responses.
We would like to find the network weights $w$ such that $p(k| x, N)$ is maximised (i.e. to make a maximum likelihood estimate given the observations):

\begin{equation}
    \max_{w}[p(k|x, w)] = \max_{w}[\text{Bin}(k|f^w(x), N)]
\end{equation}
\begin{equation}
     = \max_{w}[\log {N \choose k} + k \log {f^{w}(x)} + (N - k) \log (1 - f^{w}(x))]
\end{equation}

The combinatorial term is fixed and hence our objective function to minimise is
\begin{equation}
    \mathcal{L} = k \log {f^{w}(x)} + (N - k) \log (1 - f^{w}(x))
    \label{binomial_loss_vf}
\end{equation}

We can create a probabilistic model for $k$ by optimising our network to make maximum likelihood estimates $\hat{\rho} = f^{w}(x)$ for the latent parameter $\rho$ from which $k$ is drawn.

In short, each network $w$ predicts the response probability $\rho$ that a random volunteer will select a given answer for a given image.

\subsection{From Probabilistic to Bayesian CNN}
\label{monte_carlo_dropout}

So far, our model is probabilistic (i.e. the output is the parameter of a probabilistic model) but not Bayesian. 
If we asked $N$ volunteers, we would predict $k$ answers with a posterior of $p(k|w) = \text{Bin}(k|f^w(x), N)$ (where $f^w(x)$ is our network prediction of $\rho$ for galaxy $x$).
However, this treats the model, $w$, as fixed and known.
Instead, the Bayesian approach treats the model itself as a random variable.

Intuitively, there are many possible models that could be trained from the same training data $\mathcal{D}$.
To predict the posterior of $k$ given $\mathcal{D}$, we should marginalise over these possible models:

\begin{equation}
    \label{marginalise_over_models}
    p(k|x, \mathcal{D}) = \int p(k|x, w) p (w|\mathcal{D}) dw
\end{equation}

We need to know how likely we were to train a particular model $w$ given the data available, $p(w|\mathcal{D})$.
Unfortunately, we don't know how likely each model is. We only observe the single model we actually trained.

Instead, consider dropout \citep{Srivastava2014}. 
Dropout is a regularization method that temporarily removes random neurons according to a Bernoulli distribution, where the probability of removal (`dropout rate') is a hyperparameter to be chosen.
Dropout may be interpreted as taking the trained model and permuting it into a different one \citep{Srivastava2014}.
\cite{Gal2016Uncertainty} introduced the approach of approximating the distributions of models one might have trained, but didn't, with the distribution of networks from applying dropout:

\begin{equation}
    \label{dropout_approximation}
    p(w|\mathcal{D}) \approx q^{\ast}
\end{equation}
removing neurons according to dropout distribution $q^{\ast}$. 
This is the Monte Carlo Dropout approximation (hereafter MC Dropout).
See Appendix B for a more formal overview.

Choosing the dropout rate affects the approximation; greater dropout rates lead the model to estimate higher uncertainties (on average).
Following convention, we arbitrarily choose a dropout rate of 0.5.
We discuss the implications of using an arbitrary dropout rate, and opportunities for improvement, in Section \ref{discussion}.

Applying MC Dropout to marginalise over models (Eqn. \ref{marginalise_over_models}):
\begin{equation}
    p(k|x, \mathcal{D}) = \int p(k|x, w) q^{\ast} dw
\end{equation}

In practice, following \cite{Gal2016Uncertainty}, we sample from $q^{\ast}$ with $T$ forward passes using dropout \textit{at test time} (i.e. Monte Carlo integration):

\begin{equation}
    \int p(k|x, w) q^\ast dw \approx \frac{1}{T} \sum_{t} p(k|x, w_t)
\end{equation}

Using MC Dropout, we can improve our posteriors by (approximately) marginalising over the possible models we might have trained.

To demonstrate our probabilistic model and the use of MC Dropout, we train models to predict volunteer responses to the `Smooth or Featured' and `Bar' questions on Galaxy Zoo 2 (Section \ref{application_to_gz}).

\subsection{Data - Galaxy Zoo 2}
\label{data_gz2}

Galaxy Zoo 2 (GZ2) classified all 304,122 galaxies from the Sloan Digital Sky Survey (SDSS) DR7 Main Galaxy Sample \citep{Strauss2002, Abazajian2009} with $r$ < 17 and \texttt{petroR90\_r}\footnote{\texttt{petroR90\_r} is the Petrosian radius which contains 90\% of the $r$-band flux} > 3 arcsec.
Classifying 304,122 galaxies required  $\sim$ 60 million volunteer responses collected over 14 months.

GZ2 is the largest homogenous galaxy sample with reliable measurements of detailed morphology, and hence an ideal data source for this work.
GZ2 has been extensively used as a benchmark to compare machine learning methods for classifying galaxy morphology.
The original GZ2 data release \citep{Willett2013} included comparisons with (pre-CNN) machine learning methods by \cite{Baillard2011} and \cite{Huertas-Company2010}.
GZ2 subsequently provided the data for seminal work on CNN morphology classification \citep{Dieleman2015} and continues to be used for validating new approaches \citep{Sanchez2017, Khan2018}. 

We use the `GZ2 Full Sample' catalogue (hereafter `GZ2 catalogue'), available from \href{data.galaxyzoo.org}{data.galaxyzoo.org}. 
To avoid the possibility of duplicated galaxies or varying depth imaging, we exclude the `stripe82' subset.

The GZ2 catalogue provides aggregate volunteer responses at each of the three post-processing stages: raw vote counts (and derived vote fractions), consensus vote fractions, and redshift-debiased vote fractions.
The raw vote counts are simply the number of users who selected each answer.
The consensus vote fractions are calculated by iteratively re-weighting each user based on their overall agreement with other users.
The debiased fractions estimate how the galaxy would have been classified if viewed at $z=0.03$ \citep{Hart2016}.
Unlike recent work \citep{Sanchez2017,Khan2018}, we use the raw vote counts.
The redshift-debiased fractions estimate the \textit{true} morphology of a galaxy, not what the image actually \textit{shows}.
To predict what volunteers would say about an image, we should only consider what the volunteers see. 
We believe that debiasing is better applied after predicting responses, not before.
We caution the reader that our performance metrics are therefore not directly comparable to those of \cite{Sanchez2017} and \cite{Khan2018}, who use the debiased fractions as ground truth.

\subsection{Application}
\label{application_to_gz}

\subsubsection{Tasks}

To test our probabilistic CNNs, we aim to predict volunteer responses for the `Smooth or Featured' and `Bar' questions. 

The `Smooth or Featured' question asks volunteers `Is the galaxy simply smooth and rounded, with no sign of a disk?' with (common\footnote{`Smooth or Featured' includes a third `Artifact' answer. However, artifacts are sufficiently rare (0.08\% of galaxies have `Artifact' as the majority response) that predicting `Smooth' or `Not Smooth' is sufficient to separate smooth and featured galaxies in practice}) answers `Smooth' and `Featured or Disk'.
As `Smooth or Featured' is the first decision tree question, this question is always asked, and therefore every galaxy has $\sim$ 40 `Smooth or Featured` responses\footnote{Technical limitations during GZ2 caused 26,530 galaxies to have $N < 36$. We exclude these galaxies for simplicity.}.
With $N$ fixed to $\sim 40$ responses, the loss function (Eqn. \ref{binomial_loss_vf}) depends only on $k$ (for a given model $w$).

\begin{figure}
    \centering
    \includegraphics[width=\columnwidth]{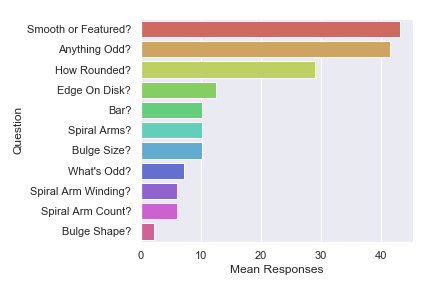}
    \caption{
        Mean responses ($N$) by GZ2 question. 
        Being the first question, `Smooth or Featured' has an unusually high ($\sim 40$) number of responses.
        Most questions (6 of 11), including `Bar', are only asked for `Featured' galaxies, and hence have only $\sim 10$ responses.
        Training CNNs while accounting for the label uncertainty caused by low $N$ responses is a key goal of this work.
        }
    \label{vote_histograms}
\end{figure}

The `Bar' question asks volunteers `Is there a sign of a bar feature through the center of the galaxy?' with answers `Bar (Yes)' and `No Bar'.
Because `Bar' is only asked if volunteers respond `Featured' and `Not Edge-On' to previous questions, each galaxy can have anywhere from 0 to 40 total responses -- typically around 10 (Figure \ref{vote_histograms}).
This scenario is common; only 2 questions are always asked, and most questions have $N << 40$ total responses (Figure \ref{vote_histograms}).
Building probabilistic CNNs that learn better by appreciating the varying count uncertainty in volunteer responses is a key advantage of our design.
We achieve this by maximising the likelihood of the observed responses given our predicted `typical' response and $N$ (Section \ref{probabilistic_prediction_with_cnn}).

\subsubsection{Architecture}

\begin{figure}
    \centering
    \includegraphics[width=\columnwidth]{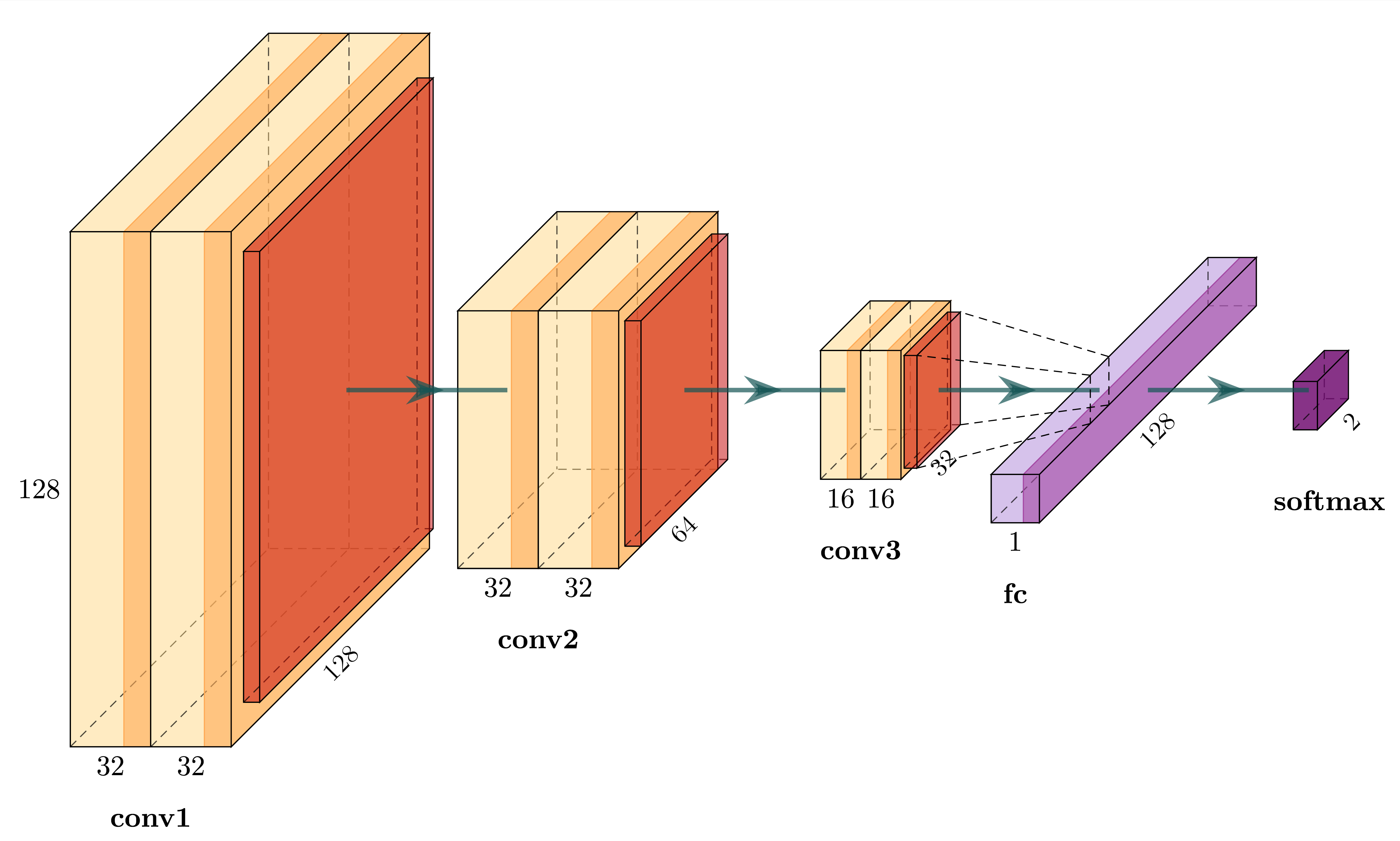}
    \caption{
        The CNN architecture used throughout. 
        The input image, after applying augmentations (Section \ref{augmentations}), is of dimension 128x128x1. 
        The first pair of convolutional layers are each of dimension 128x128x32 with 3x3 kernels. 
        We then max-pool down to a second pair of convolutional layers of dimension 64x64x32 with 3x3 kernels, then again to a final pair of dimension 32x32x16 with 3x3 kernels.
        We finish with a 128-neuron linear dense layer and a 2-neuron softmax dense layer.}
    \label{cnn_architecture}
\end{figure}

Our CNN architecture is shown in Figure \ref{cnn_architecture}.
This architecture is inspired by VGG16 \citep{Simonyan2014}, but scaled down to be shallower and narrower in order to fit our computational budget.
We use a softmax final layer to ensure the predicted typical vote fraction $\rho$ lies between 0 and 1, as required by our binomial loss function (Equation \ref{binomial_loss_vf}).

We are primarily concerned with accounting for label uncertainty and predicting posteriors, rather than maximising performance metrics.
That said, our architecture is competitive with, or outperforms, previous work (Section \ref{previous_work}).
Our overall performance can likely be significantly improved with more recent architectures \citep{Szegedy2015, He2015, Huang2017} or a larger computational budget.

\subsubsection{Augmentations}
\label{augmentations}

To generate our training and test images, we resize the original 424x424x3 pixel GZ2 \texttt{png} images shown to volunteers into 256x256x3 \texttt{uint8}\footnote{Unsigned 8-bit integer i.e. 0-255 inclusive. After rescaling, this is sufficient to express the dynamic range of the images (as judged by visual inspection) while significantly reducing memory requirements vs. the original 32-bit float flux measurements.} matrices and save these matrices in TFRecords (to facilitate rapid loading).
When serving training images to our model, each image has the following transformations applied:

\begin{enumerate}
    \item Average over channels to create a greyscale image
    \item Random horizontal and/or vertical flips
    \item Rotation through an angle randomly selected from $0\degree$ to $90\degree$ (using nearest-neighbour interpolation to fill pixels) 
    \item Adjusting the image contrast to a contrast uniformly selected from 98\% to 102\% of the original contrast
    \item Cropping either randomly (`Smooth or Featured') or centrally (`Bar') according to a zoom level uniformly selected from 1.1x to 1.3x (`Smooth or Featured') or 1.7x to 1.9x (`Bar')
    \item Resizing to a target size of 128x128(x1)
\end{enumerate}

We train on greyscale images because colour is often predictive of galaxy type (E and S0 are predominantly redder, while S are bluer, \citealt{Roberts2002}) and we wish to ensure that our classifier does not learn to make biased predictions from this correlation. 
For example, a galaxy should be classified as smooth because it appears smooth, and not because it is red and therefore more likely to be smooth.
Otherwise, we bias any later research investigating correlations between morphology and colour.

Random flips, rotations, contrast adjustment, and zooms (via crops) help the CNN learn that predictions should be invariant to these transformations - our predictions should not change because the image is flipped, for example.
We choose a higher zoom level for `Bar' because the original image radius for GZ2 was designed to show the full galaxy and any immediate neighbours \citep{Willett2013} yet bars are generally found in the center of galaxies \citep{Kruk2017}.
We know that the `Bar' classification should be invariant to all but the central region of the image, and therefore choose to sacrifice the outer regions in favour of increased resolution in the centre.
Cropping and resizing are performed last to minimise resolution loss due to aliasing.
Images are resized to match our computational budget. 

We also apply these augmentations at test time. 
This allows us to marginalise over any unlearned invariance using MC Dropout, as part of marginalising over networks (Section \ref{monte_carlo_dropout}).
Each permuted network makes predictions on a uniquely-augmented image.
The aggregated posterior (over many forward passes $T$) is therefore independent of e.g. orientation, enforcing our domain knowledge.

\subsection{Experimental Setup}
\label{experimental_setup}

For each question, we randomly select 2500 galaxies as a test subset and train on the remaining galaxies (following the selection criteria described in Section \ref{data_gz2}).
Unlike \cite{Sanchez2017} and \cite{Khan2018}, we do not select a `clean' sample of galaxies with extreme vote fractions on which to train.
Instead, we take full advantage of the responses collected for every galaxy by carefully accounting for the vote uncertainty in galaxies with fewer responses (Eqn \ref{binomial_loss_vf}).

For `Smooth or Featured', we use a final training sample of 176,328 galaxies.
For `Bar', we train and test only on galaxies with $N_{\text{bar}} \geq 10$ (56,048 galaxies).
Without applying this cut, we find that models fail to learn; performance fails to improve from random initialisation.
This may be because galaxies with $N_{\text{bar}} < 10$ must have $k_{\text{featured}} < 10$ and so are almost all smooth and unbarred, leading to increasingly unbalanced typical vote fractions $\rho$.

Training was performed on an Amazon Web Services (AWS) \texttt{p2.xlarge} EC2 instance with an NVIDIA K80 GPU. 
Training each model from random initialisation takes approximately eight hours.

Using the trained models, we make predictions $\hat{\rho}$ for the typical vote fraction $\rho$ of each galaxy in the test subsets.
We then evaluate performance by comparing $p(k|\hat{\rho}, N)$, our posterior for $k$ positive responses from $N$ volunteers, with the observed $k$ from the $N$ Galaxy Zoo volunteers asked.

\subsection{Results}
\label{probabilistic_results}

We find that our probabilistic CNNs produces posteriors which are reliable and informative. 

\begin{figure*}
    \centering
    \includegraphics[width=0.6\textwidth]{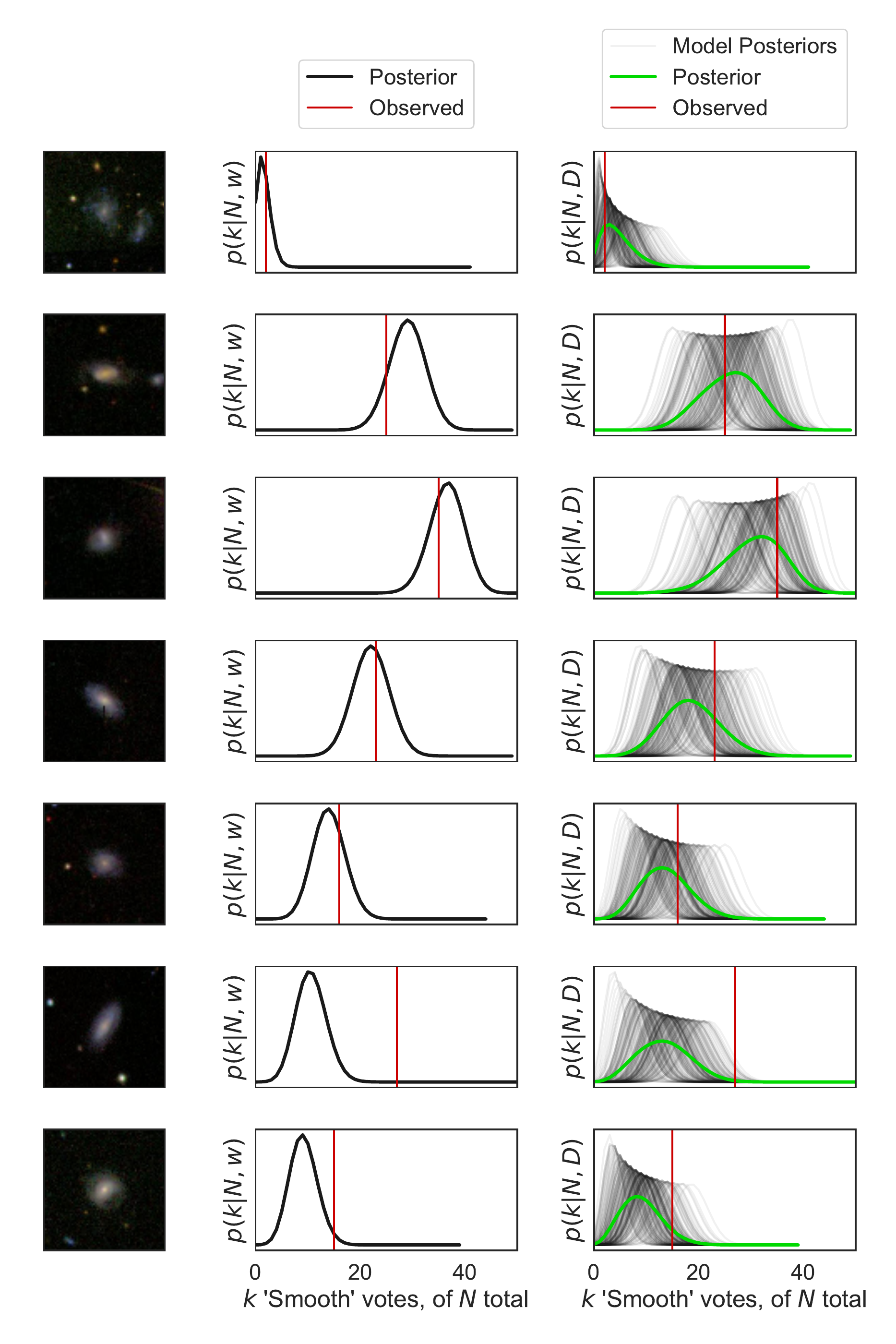}
    \caption{
        Posteriors for $k$ of $N$ volunteers answering `Smooth' to the question `Smooth or Featured?'). 
        Each row is a randomly selected galaxy.  
        Overplotted in red is the actual $k$ measured from $N \sim 40$ volunteers. 
        The left column shows the galaxy in question, as presented to the network (following the augmentations described in Section \ref{augmentations}).
        The central column shows the posterior predicted by a single network (black), while the right column shows the posterior marginalised (averaged) over 30 MC-dropout-approximated `networks' (green) as well as from each `network' (grey).
        While the posterior from a single network is fixed to a binomial form, the marginalised posteriors from many `networks' can take any form.
        The posterior from a single network is generally more confident (narrower); we later show that a single network is overconfident, and many `networks' are better calibrated.
        }
    \label{dual_posteriors_smooth}
\end{figure*}

\begin{figure*}
    \centering
    \includegraphics[width=0.6\textwidth]{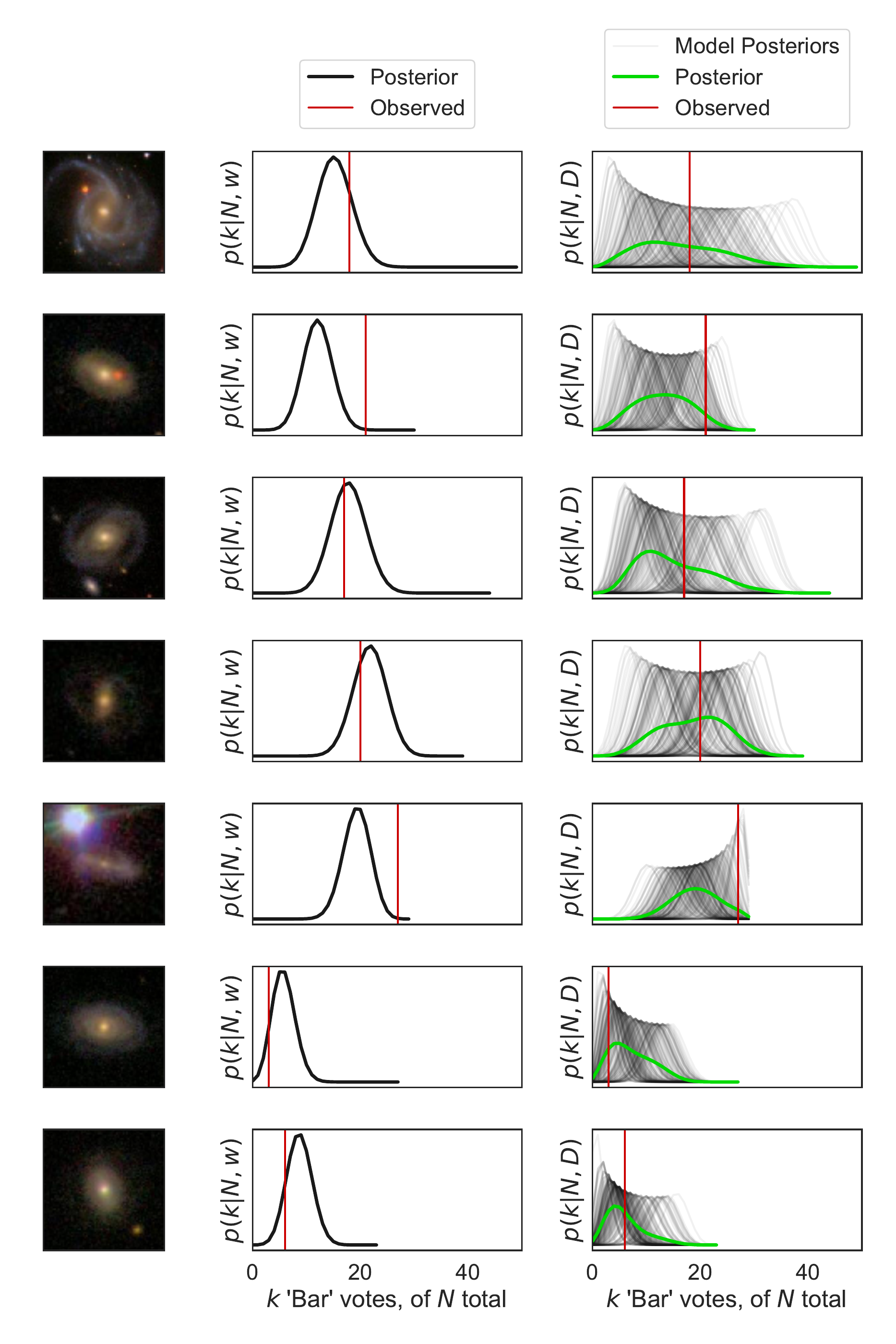}
    \caption{
        As for Figure \ref{dual_posteriors_smooth}, but showing posteriors for $k$ of $N$ volunteers answering `Bar (Yes)' to the question `Bar?'.
        Unlike `Smooth or Featured', $N$ varies significantly between galaxies, and hence so does the spread (uncertainty in $k$) and absolute width (highest possible $k$) of the posterior.
        }
    \label{dual_posteriors_bar}
\end{figure*}

For each question, we first compare a random selection of posteriors from either 1 or 30 MC Dropout forward passes (i.e. 1 or 30 MC-dropout-approximated `networks').
Figures \ref{dual_posteriors_smooth} and \ref{dual_posteriors_bar} show our posteriors for `Smooth or Featured' and `Bar', respectively.

\textit{Without} MC Dropout, our posteriors are binomial. 
The spread of each posterior reflects two effects.
First, the spread reflects the extremity of $\hat{\rho}$ that previous authors have expressed as `volunteer agreement' or `confidence' \citep{Dieleman2015,Sanchez2017}.
Bin$(k|\hat{\rho}, N)$ is narrower where $\hat{\rho}$ is close to 0 or 1.
Second, the spread reflects $N$, the number of volunteers asked. 
For `Smooth or Featured', where $N$ is approximately fixed, this second effect is minor. 
For `Bar', where $N$ varies significantly between 10 and $\sim$ 40, the posteriors are more spread (less precise) where fewer volunteers have been asked.

\textit{With} MC Dropout, our posteriors are a superposition of Binomials from each forward pass, each centered on a different $\hat{\rho_t}$.
In consequence, the MC Dropout posteriors are more uncertain. 
This matches our intuition - by marginalising over the different weights and augmentations we might have used, we expect our predictions to broaden.

Given that each single network is relatively confident and the MC-dropout-marginalised model is relatively uncertain, which should be used?
We prefer posteriors which are well-calibrated i.e. which reflect the true uncertainty in our predictions.

To quantify calibration, we introduce a novel method; we compare the predicted and observed vote fractions $\frac{k}{N}$ within increasing ranges of acceptable error. We outline this procedure below.

Choose some maximum acceptable error $\epsilon$ in predicting each vote fraction $v = \frac{k}{N}$.
Over all galaxies, sum the total probability (from our predicted posteriors) that $v_i = \hat{v_i} \pm \epsilon$ for each galaxy $i$.
We call this the expected count: how many galaxies the posterior suggests should have $v$ within $\epsilon$ of the model prediction $\hat{v}$.
For example, our `Bar' model expects 2320 of 2500 galaxies in the `Bar' test set to have an observed $v$ within $\pm 0.20$ of $\hat{v}$.

\begin{equation}
    C_\text{expected} = \sum_{i}^{N_\text{galaxies}} \sum_{j > \hat{k} - N\epsilon}^{j < \hat{k} + N\epsilon} p(j | \hat{\rho_i}, N_i)    
\end{equation}

Next, over all galaxies, count how often $v_i$ is within that maximum error $v_i = \hat{v_i} \pm \epsilon$.
We call this the `actual' count: how many galaxies are actually observed to have $v_i$ within $\epsilon$ of the model prediction $\hat{v_i}$.
For example, we observe 2075 of 2500 galaxies in the `Bar' test set to have $v_i$ within $\pm 0.20$ of $\hat{v}$.

\begin{equation}
    C_\text{actual} = \sum_{i}^{N_\text{galaxies}} \sum_{j > \hat{k_i} - N\epsilon}^{j < \hat{k_i} + N\epsilon} \delta(k_i - j)
\end{equation}

For a perfectly calibrated posterior, the actual and expected counts would be identical: the model would be correct (within some given maximum error) as often as it expects to be correct.
For an overconfident posterior, the expected count will be higher, and for an underconfident posterior, the actual count will be higher.

\begin{figure}
    \begin{subfigure}[b]{\columnwidth}
        \includegraphics[width=\columnwidth]{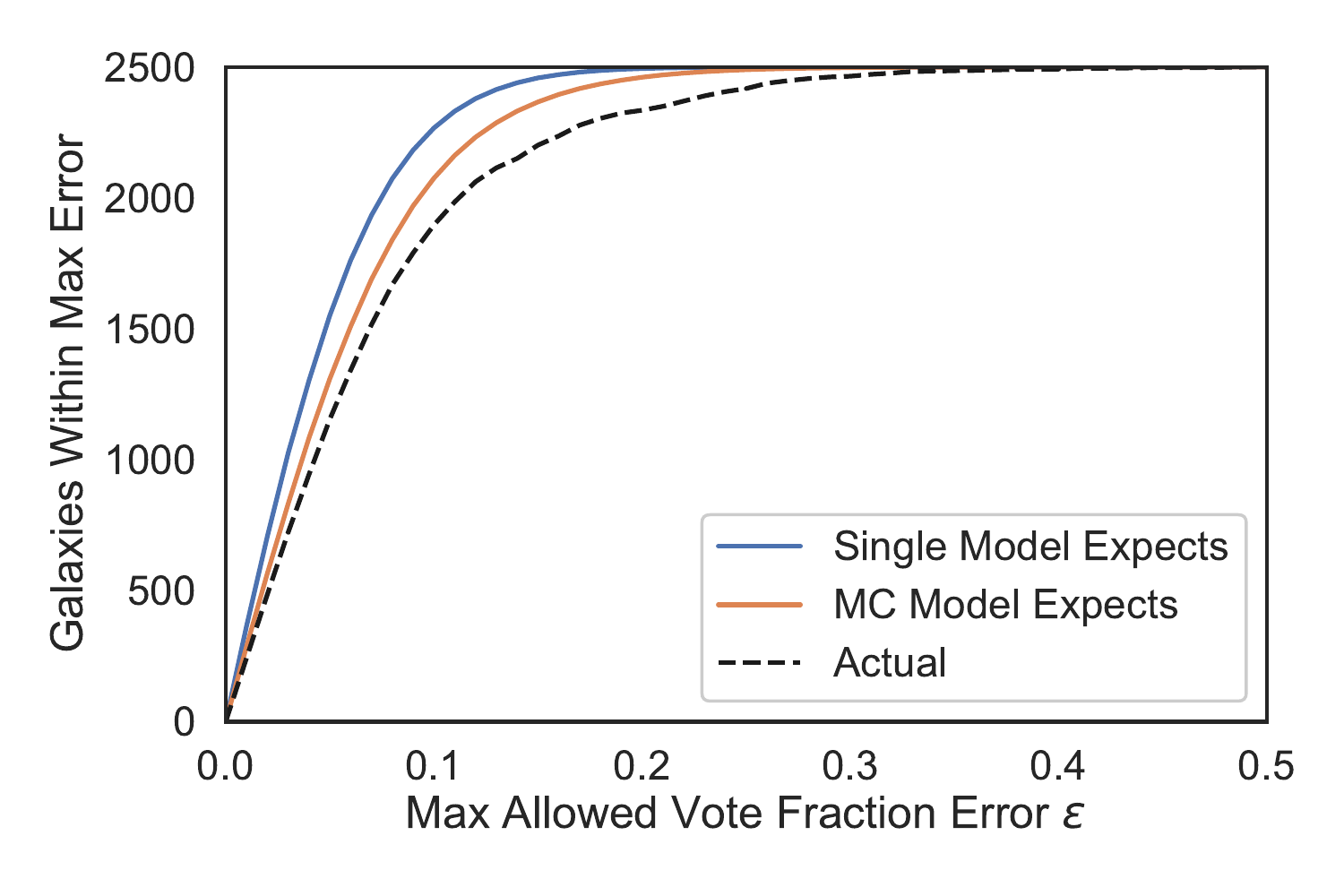}
        \caption{Calibration for `Smooth or Featured'}
    \end{subfigure}
    \begin{subfigure}[b]{\columnwidth}
        \includegraphics[width=\columnwidth]{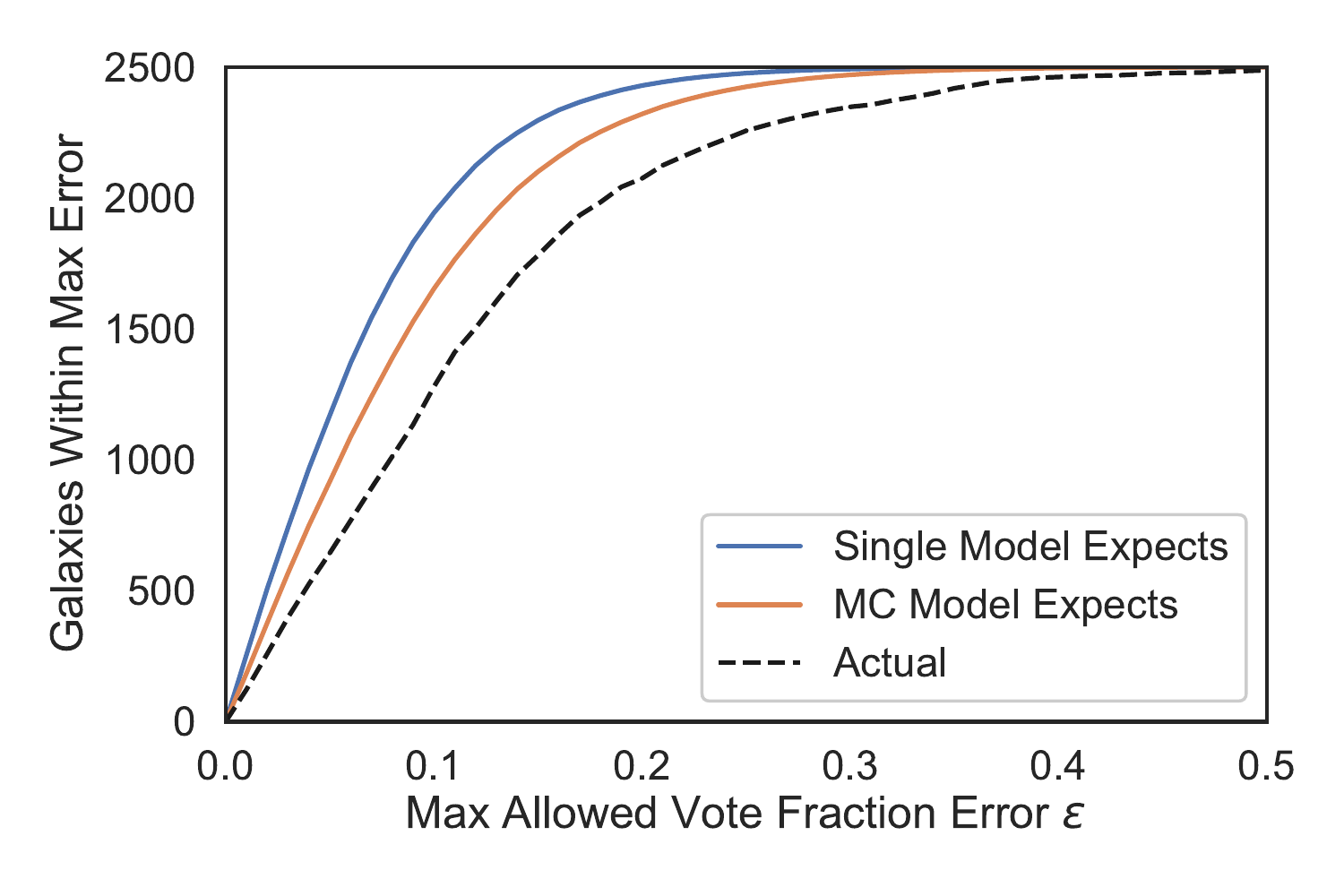}
        \caption{Calibration for `Bar'}
    \end{subfigure}
    \caption{Calibration of CNN-predicted posteriors, showing the expected vs. actual count of galaxies within each acceptable maximum vote fraction error range ($\epsilon$).
     Our probabilistic model is fairly well-calibrated (similar expected and actual counts), with a significant improvement from applying MC Dropout.}
    \label{coverage_comparison}
\end{figure}

We find that our predicted posteriors of volunteer votes are fairly well-calibrated; our model is correct approximately as often as it \textit{expects} to be correct.
Figure \ref{coverage_comparison} compares the expected and actual counts for our model, choosing $\epsilon$ between 0 and 0.5.
Tables \ref{tab:smooth_calibration} and \ref{tab:bar_calibration} show calibration results for our `Smooth' and `Bar' models, with and without MC Dropout, evaluated on their respective test sets.
Coverage error is calculated as:

\begin{equation}
    \text{Coverage error} = \frac{C_\text{expected} - C_\text{actual}}{C_\text{actual}}.
\end{equation}

\begin{table}
    \centering
    \caption{Calibration results for predicting the probability that $v \pm \epsilon$ fraction of volunteers respond `Smooth', with and without applying MC Dropout.}
    \label{tab:smooth_calibration}
    \begin{tabular}{lcc}
        \hline
        Max Error $\epsilon$ & Coverage Error without MC & Coverage Error with MC\\
        \hline
        0.02 & 49.6\% & 16.5\%\\
        0.05 & 38.5\% & 13.4\%\\
        0.10 & 26.1\%\ & 9.4\%\\
        0.20 & 7.9\% & 5.4\%\\
        \hline
    \end{tabular}
\end{table}

\begin{table}
    \centering
    \caption{Calibration results for predicting the probability that $v \pm \epsilon$ fraction of volunteers respond `Bar', with and without applying MC Dropout.}
    \label{tab:bar_calibration}
    \begin{tabular}{lcc}
        \hline
        Max Error $\epsilon$ & Coverage Error without MC & Coverage Error with MC  \\
        \hline
        0.02 & 92.2\% & 45.5\%\\
        0.05 & 85.5\% & 42.4\%\\
        0.10 & 57.8\% & 29.2\%\\
        0.20 & 22.6\% & 11.8\%\\
        \hline
    \end{tabular}
\end{table}

For both questions, the single network (without using MC Dropout) is visibly overconfident.
The MC-dropout-marginalised network shows a significant improvement in calibration over the single network.
We interpret this as evidence for the importance of marginalising over both networks and augmentations in accurately estimating uncertainty (Section \ref{monte_carlo_dropout}).

When making precise predictions, the MC-dropout-marginalised network remains somewhat overconfident.
However, as the acceptable error $\epsilon$ is allowed to increase, the network is increasingly well-calibrated. 
For example, the predicted probability that $v \pm 0.02$ (i.e. $\epsilon = 0.02$) $k$ of $N$ volunteers respond `Bar' is over-estimated by $\sim$ 45\%.
In contrast, the predicted probability that $k \pm 0.2$ (i.e. $\epsilon = 0.2$) of $N$ volunteers respond `Bar' is $\sim$ 10\% of the true probability.
We discuss future approaches to further improve calibration in Section \ref{discussion}.

\begin{figure}
    \begin{subfigure}[b]{\columnwidth}
        \includegraphics[width=\columnwidth]{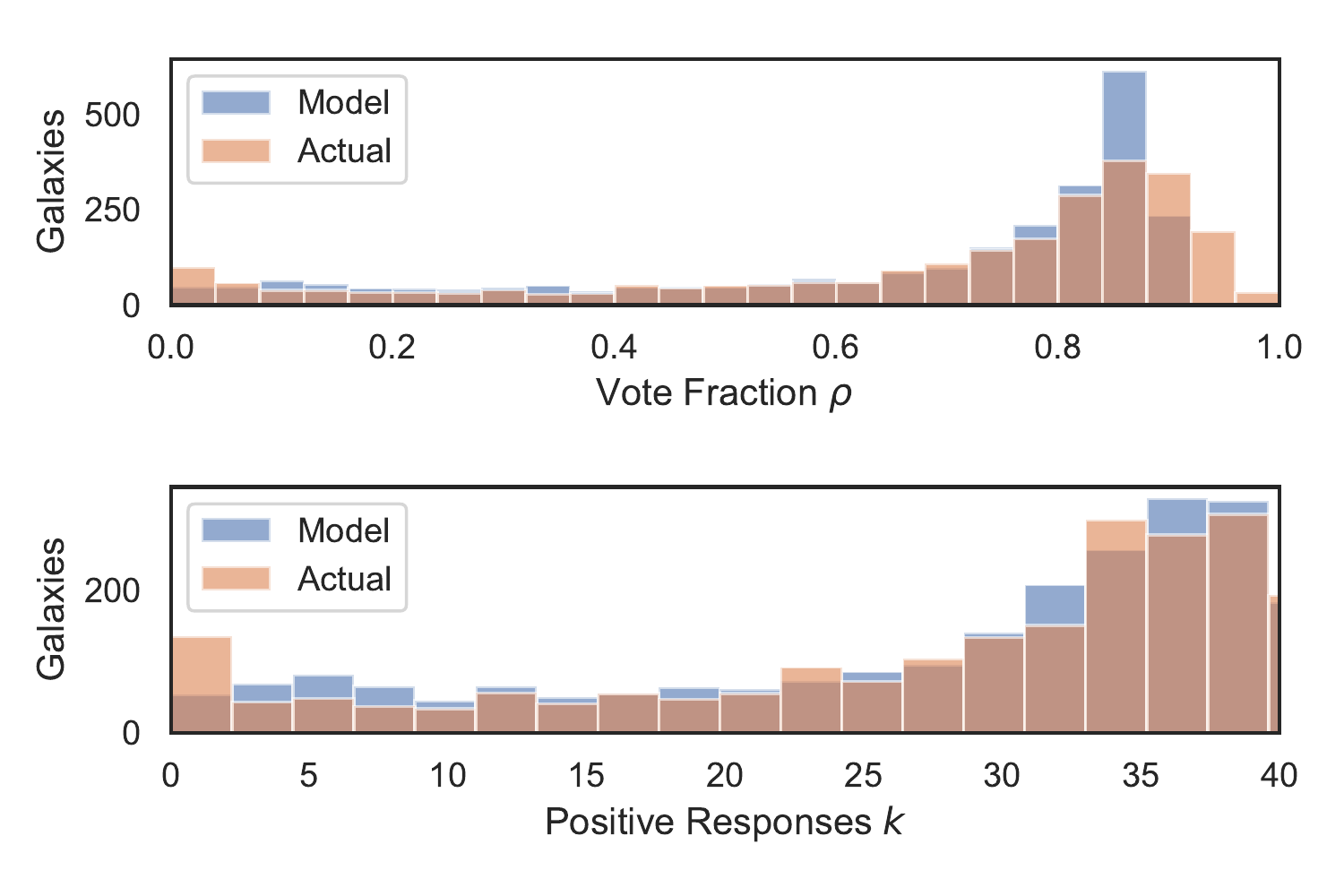}
        \caption{Distribution of $k$ and $\rho$ for `Smooth or Featured'}
    \end{subfigure}
    \begin{subfigure}[b]{\columnwidth}
        \includegraphics[width=\columnwidth]{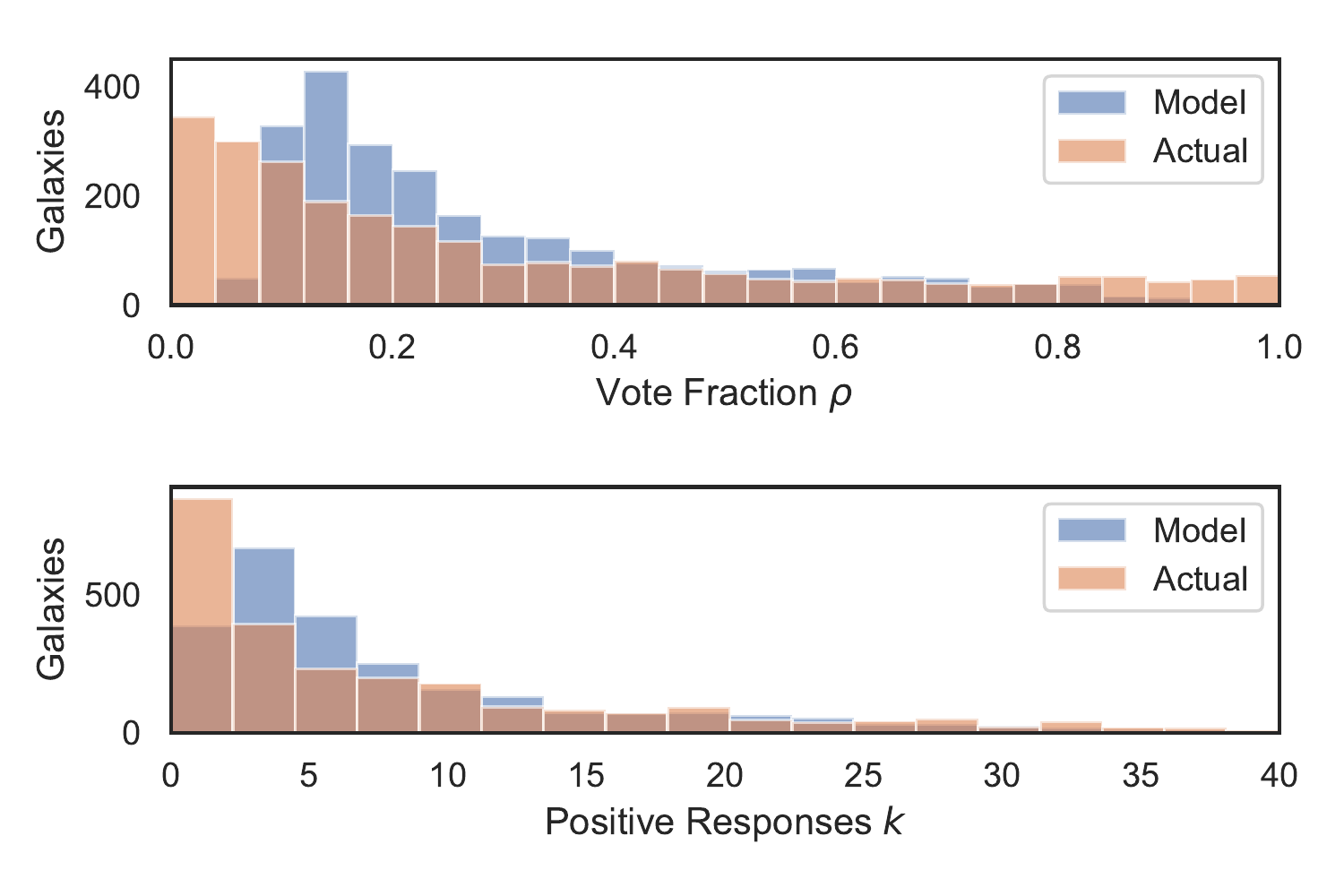}
        \caption{Distribution of $k$ and $\rho$ for `Bar'}
    \end{subfigure}
    \caption{Comparison between the distribution of predicted or observed $\rho$ and $k$ over all galaxies, for each question. 
    Upper: comparison for `Smooth or Featured'. Lower: comparison for `Bar'.
    The observed $\rho$ is approximated as  $\rho_{\text{proxy}} = \frac{k}{N}$
    The distributions of predicted $\rho$ and $k$ closely match the observed distributions, indicating our models are globally unbiased.
    The only significant deviation is near extreme $\rho$ and $k$, which our models are `reluctant' to predict.}
    \label{posterior_over_all_galaxies}
\end{figure}

A key method for galaxy evolution research is to compare the distribution of some morphology parameter across different samples (e.g. are spirals more common in dense environments, \citealt{Wang2018}, do bars fuel AGN, \citealt{Galloway2015}, do mergers inhibit LERGs, \citealt{Gordon2019}, etc.)
We would therefore like the distribution of predicted $\hat{\rho}$ and $\hat{k}$, over all galaxies, to approximate the observed distribution of $\rho$\footnote{The `observed' $\rho$ is approximated as  $\rho_{\text{proxy}} = \frac{k}{N}$, which has a similar distribution to the true (latent, unobserved) $\rho$ over a large sample.} and $k$.
In short, we would like our predictions to be \textit{globally unbiased}.
Figure \ref{posterior_over_all_galaxies} compares our predicted and actual distributions of $\rho$ and $k$.
We find that our predicted distributions for $\rho$ and $k$ match well with the observed distributions for most values of $\rho$ and $k$.
Our model appears somewhat reticent to predict extreme $\rho$ (and therefore extreme $k$) for both questions.
This may be a consequence of the difficulty in predicting the behaviour of single volunteers.
We discuss this further in Section \ref{discussion}.

Reliable research conclusions also require that model performance should not depend strongly on non-morphological galaxy parameters (mass, colour, etc).
For example, if a researcher would like to investigate correlations between galaxy mass and bars, it is important that our model is equally able to recognise
bars in high-mass and low-mass galaxies.
To check if our model is sensitive to non-morphological parameters, we use an Explainable Boosting Machine (EBM) model \citep{Lou2012,Caruana2015}.
EBM aim to predict a target variable based on tabular features by separating the impact of those features into single (or, optionally, pairwise) effects on the target variable.
They are a specific\footnote{https://github.com/microsoft/interpret} implementation of Generalised Additive Models (GAM, \citealt{Hastie2017}). GAM are of the form:

\begin{equation}
    g(y) = f_1(x) + ... + f_n(x_n)
\end{equation}
where g is identity for regression problems and $f_i$ is any learnable function.
For EBM, each $f_i$ is learned using gradient boosting with bagging of shallow regression trees.
They aim to answer the question `What is the effect on the target variable of \textit{this particular feature alone}?'
We train an EBM to predict the surprise\footnote{Recall that we quantify surprise as the likelihood of our prediction 
given the observed votes $\frac{k}{N}$ (Eqn \ref{model_likelihood}). } of our `Bar' model when making test set predictions (Section \ref{experimental_setup}),
using the human-reported morphologies and key non-morphological parameters reported in the NASA Sloan Atlas (v1.01, \citealt{Albareti2017}). 

\begin{figure}
    \centering
    \includegraphics[width=\columnwidth]{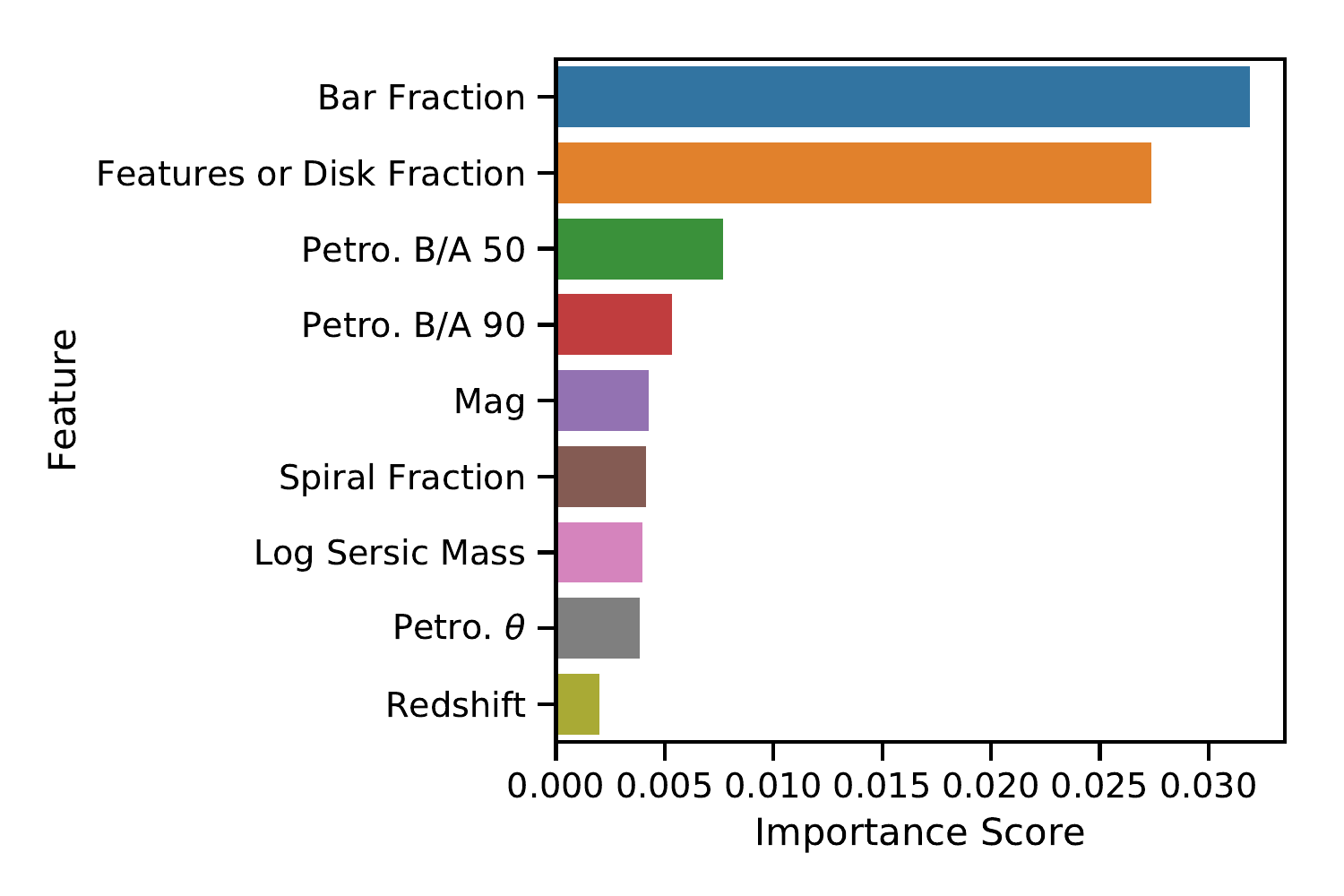}
    \caption{
        Relative importance of morphological (Features or Disk, Bar, Spiral) and non-morphological (Petro B/A, Mag, etc.) features for BCNN performance.
        Morphology fractions are the (human-reported) $\frac{k}{N}$ values from Galaxy Zoo 2.
        Petro B/A 50 and Petro B/A 90 measure the axial ratios at 50\% and 90\% of the half-light radius.
        Mag is the estimated B magnitude.
        Sersic mass is the approximate stellar mass, estimated from the single-component Sersic fit flux.
        Petro $\theta$ is the ($r$-band) Petrosian radius.
        Redshift is measured spectroscopically.
        The effect of each component is additive and independent; for example, the measured effect of spiral features does not include the effect of being featured in general.
        BCNN performance varies much less from the effect of non-morphological features than from morphological features.
        }
    \label{feature_importance}
\end{figure}

The interested reader can find our full investigation at \href{walmsley.dev/2019/bias}{www.walmsley.dev/2019/bias}, recorded as a Jupyter Notebook.
Figure \ref{feature_importance} shows the key result; the relative importance of each feature on BCNN model surprise.
We find that performance variation with respect to non-morphological parameters is much smaller than variation with respect to morphology.
Our network performs better on smooth galaxies and unbarred galaxies (plausibly because there are more training examples of such galaxies to learn from).
Inclination is the non-morphological parameter with the strongest effect on performance, and this effect is approx. 3.5-4x weaker than the effect of either smoothness or barredness above.
We are therefore confident that our model introduces no new major biases with respect to key non-morphological parameters.

\subsubsection{Comparison to Previous Work}
\label{previous_work} 

The key goals of this paper are to introduce probabilistic predictions for votes and (in the following section) to apply this to perform active learning.
However, by reducing our probabilistic predictions to point estimates, we can also provide conventional predictions and performance metrics.

Previous work has focused on deterministic predictions of either the votes \citep{Dieleman2015} or the majority response \citep{Sanchez2017, Khan2018}.
While differences in sample selection and training data prevent a precise comparison, our model performs well at both tasks.

\begin{figure}
    \begin{subfigure}[b]{\columnwidth}
        \includegraphics[width=\columnwidth]{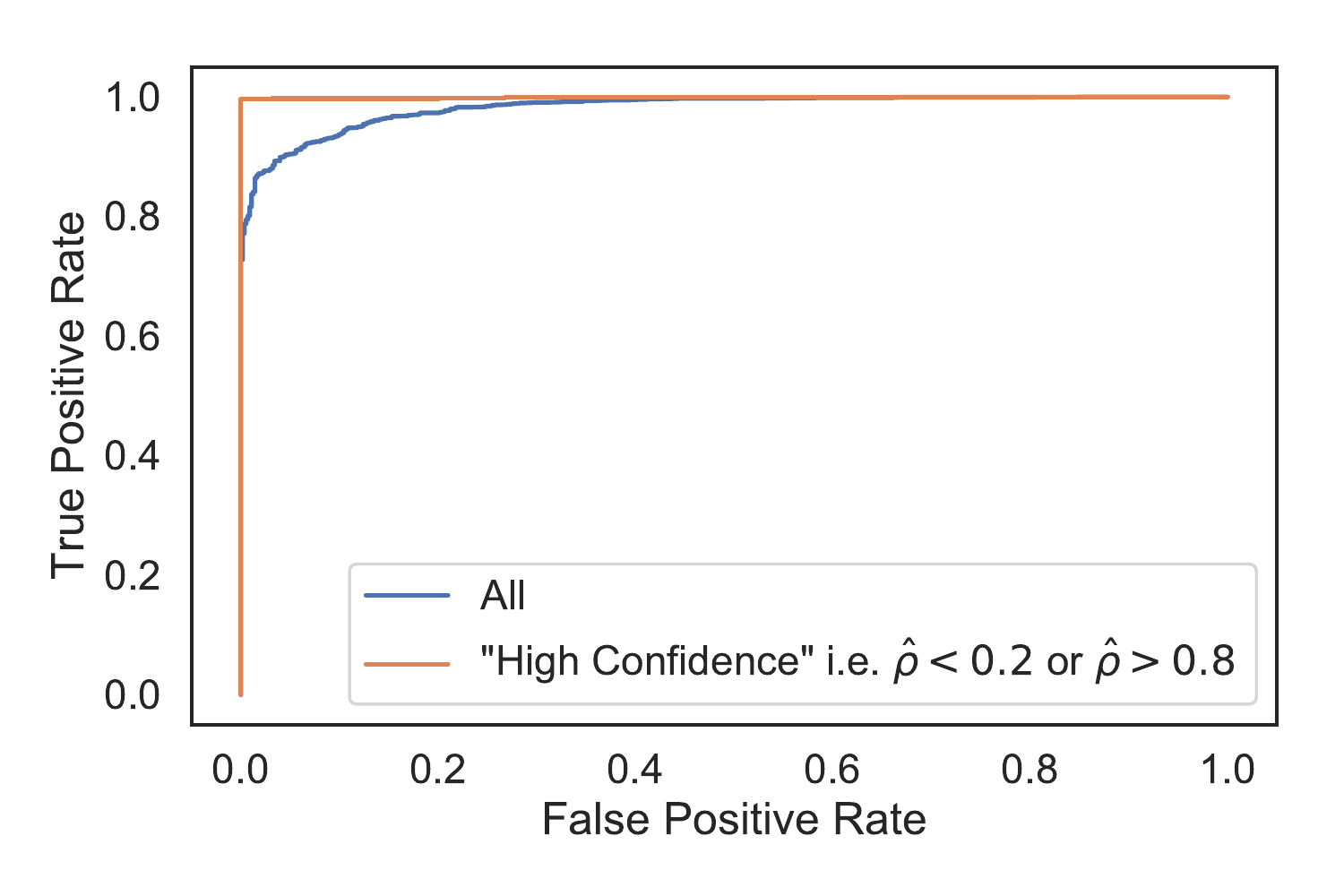}
        \caption{ROC curve for the `Smooth or Featured' question.}
    \end{subfigure}
    \begin{subfigure}[b]{\columnwidth}
        \includegraphics[width=\textwidth]{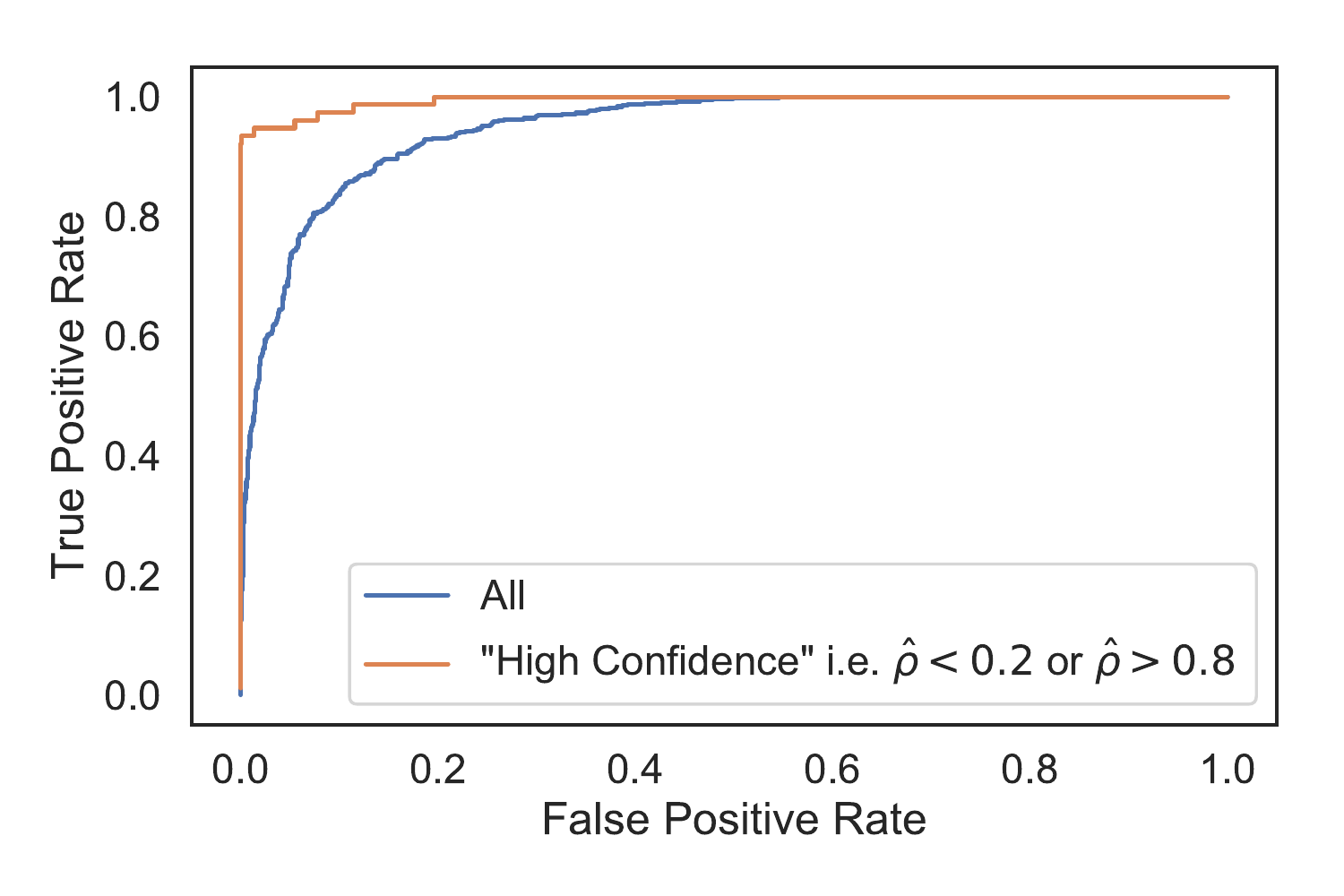}
        \caption{ROC curve for the `Bar' question.}
    \end{subfigure}
    \caption{
        ROC curves for the `Smooth or Featured' (above) and `Bar' (below) questions, as predicted by our probabilistic model.
        To generate scalar class predictions on which to threshold, we reduce our posteriors to mean vote fractions.
        For comparison to DS+18, we also include ROC curves of the subsample they describe as `high confidence' -- galaxies where the class probability (for us, $\hat{\rho}$) is extreme (1420 galaxies for `Smooth', 1174 for `Bar')}
    \label{roc_curves}
\end{figure}

\begin{figure}
    \hspace{1.75cm} \textbf{Full Sample} \hspace{1.7cm} \textbf{`High Confidence'} \hspace{1cm}
    \newline
    \centering
    \includegraphics[width=0.2\textwidth]{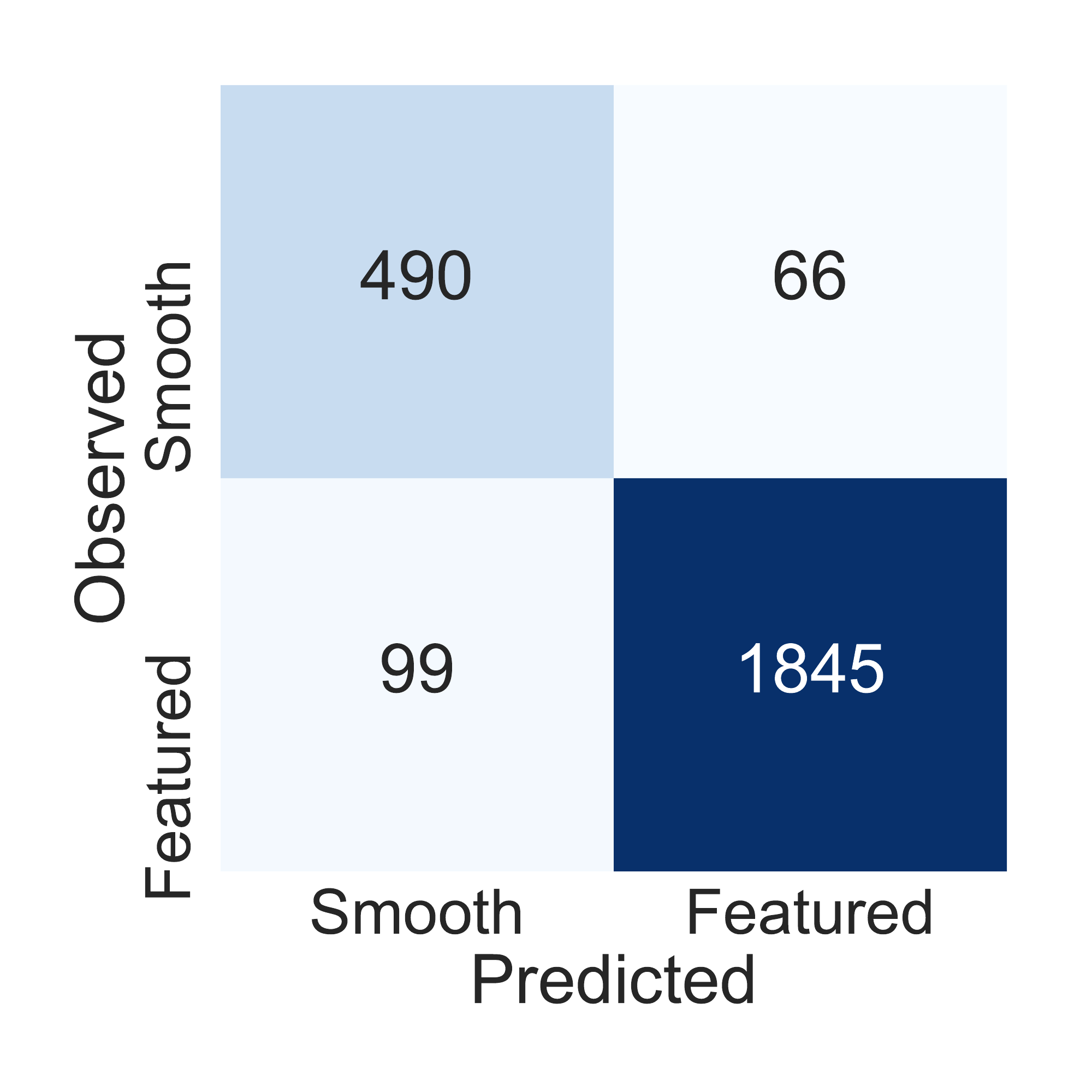}
    \includegraphics[width=0.2\textwidth]{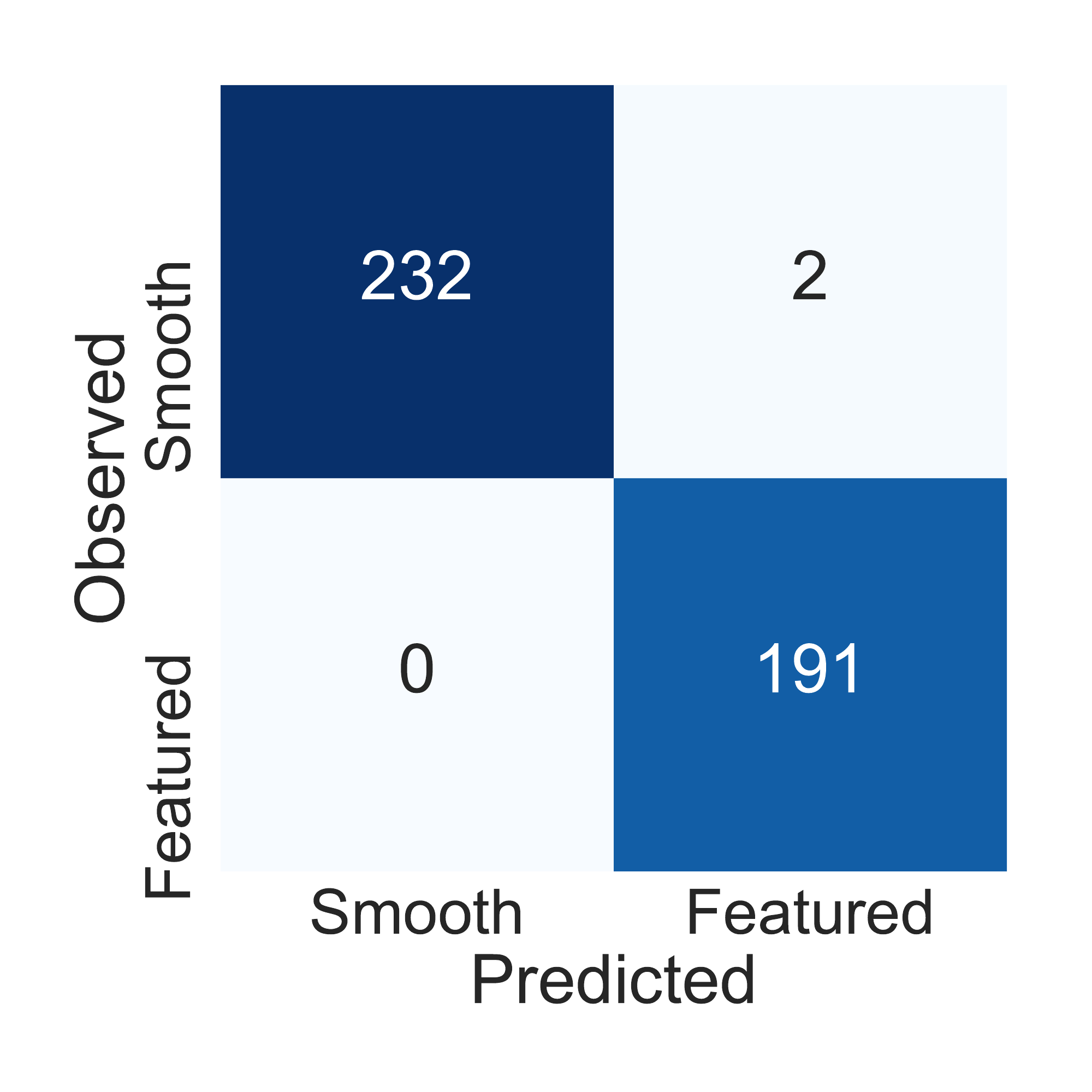}
    \includegraphics[width=0.2\textwidth]{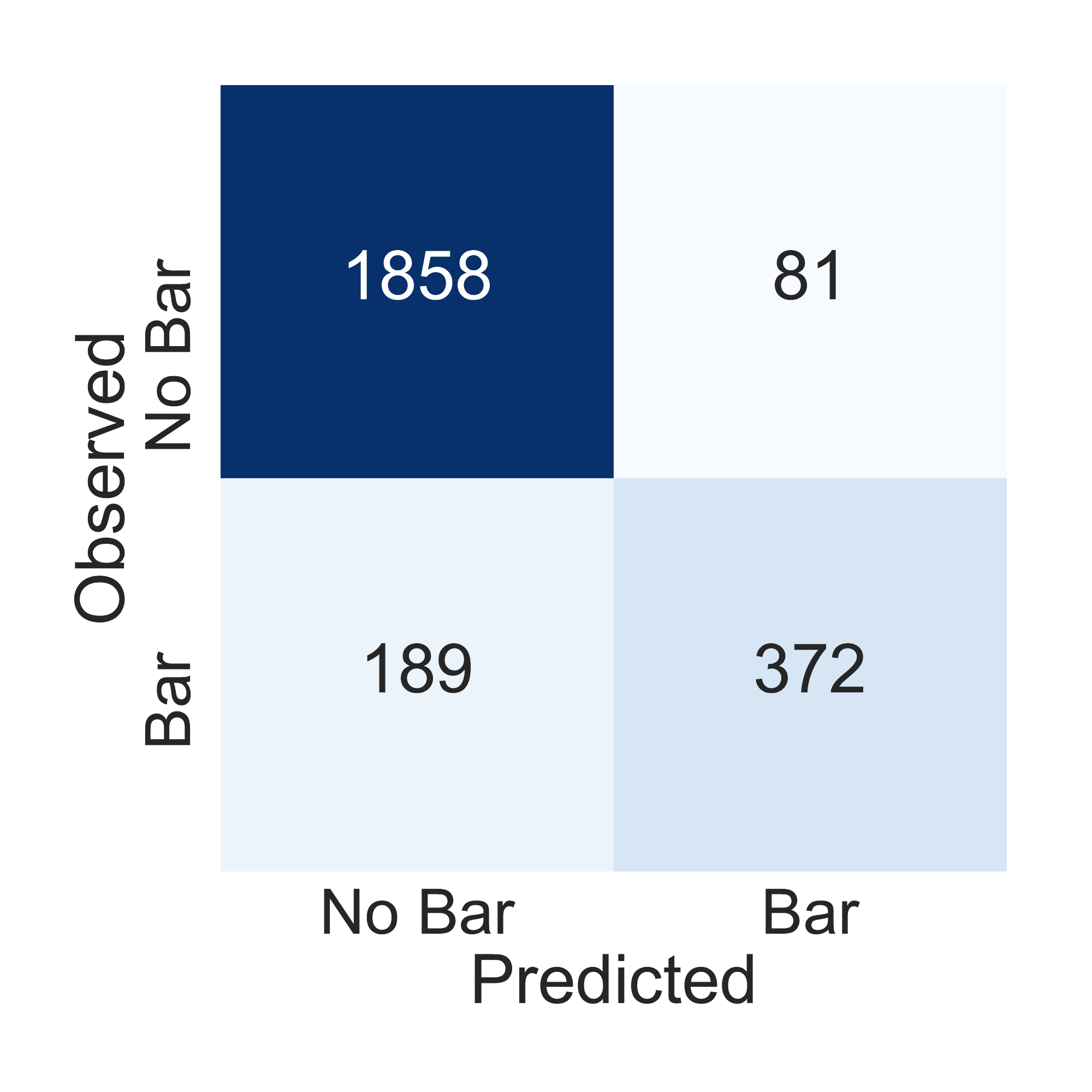}
    \includegraphics[width=0.2\textwidth]{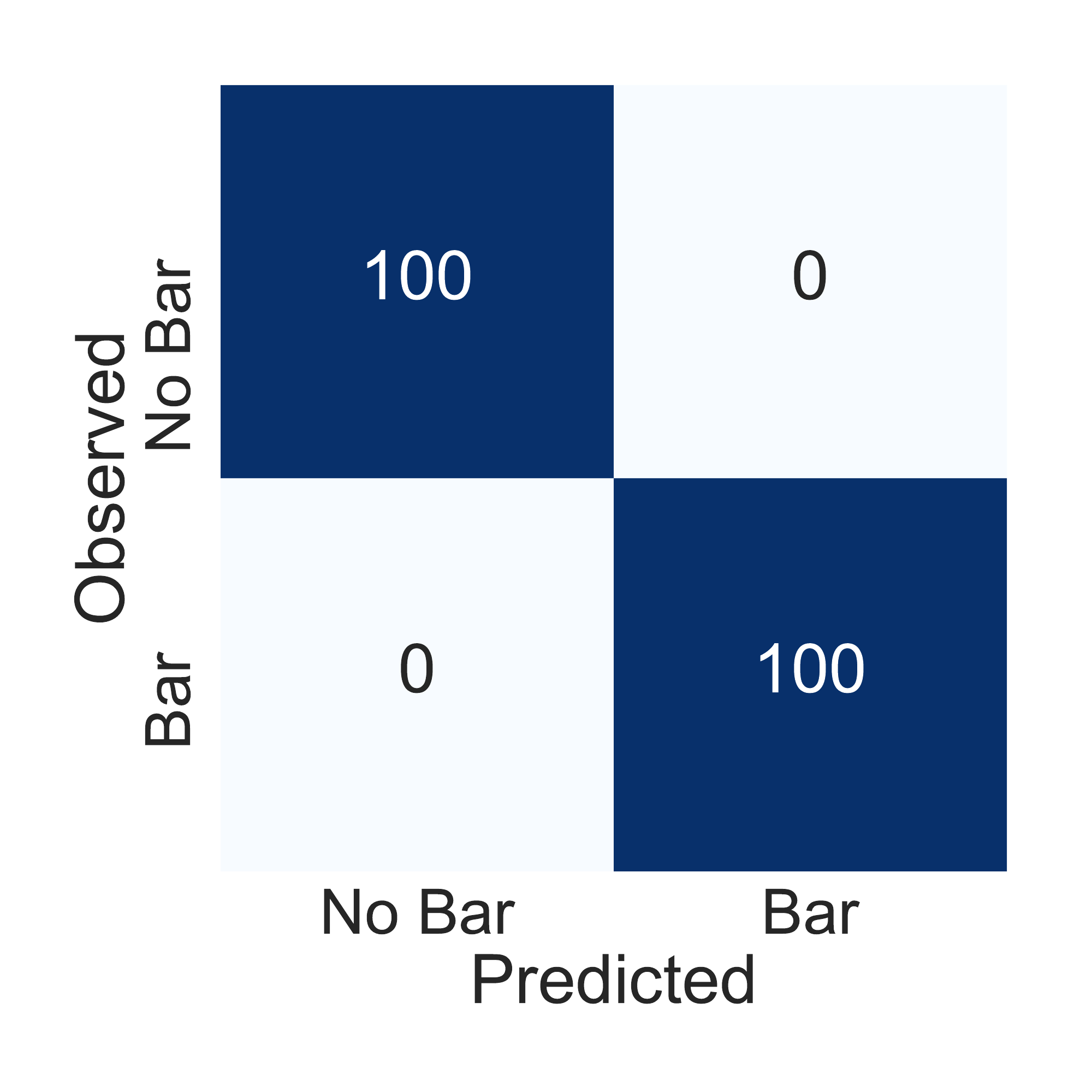}
    \caption{Confusion matrices for `Smooth or Featured' (upper row) and `Bar' (lower row) questions.
    For comparison to K+18, we also include confusion matrices for the most confident predictions (right column)
    Following K+18, we include the most confident $\sim 7.7\%$ of spirals and $\sim 9.3\%$ of ellipticals (upper right).
    Of the two galaxies where humans select `Smooth' $(\frac{k}{N} > 0.5)$ and the model selects `Featured' $(\hat{\rho} < 0.5)$, one is an ongoing smooth/featured major merger and one is smooth with an imaging artifact.
    Generalising (K+18 do not consider bars), we also show the most confident $\sim 8\%$ of barred and unbarred galaxies. 
    We achieve perfect classification for `Bar'.}
    \label{confusion_matrices}
\end{figure}

When reducing our posteriors to the most likely vote count $\hat{k}$, we achieve a root-mean-square error of 0.10 (approx. $\pm 3$ votes) for `Smooth or Featured' and 0.15 for `Bar'.
We can also reduce the same posteriors to the most likely majority responses. 
Below, we present our results in the style of the ROC curves in \cite{Sanchez2017} (hereafter DS+18, Figure \ref{roc_curves}) and the confusion matrices in \cite{Khan2018} (hereafter K+18, Figure \ref{confusion_matrices}) using our reduced posteriors.
We find that our model likely outperforms \cite{Sanchez2017} and is likely comparable with \cite{Khan2018}.

Overall, these conventional metrics demonstrate that our models are sufficiently accurate for practical use in galaxy evolution research even when reduced to point estimates.

\section{Active Learning}
\label{methods_active_learning}

In the first half of this paper, we presented Bayesian CNNs that predict posteriors for the morphology of each galaxy.
In the second, we show how we can use these posteriors to select the most informative galaxies for labelling by volunteers, helping humans and algorithms work together to do better science than either alone.

CNNs, and other deep learning methods, rely on vast training sets of labelled examples \citep{Simonyan2014, Szegedy2015, Russakovsky2015, He2015, Huang2017}.
As we argued in Section \ref{introduction}, we urgently need methods to reduce this demand for labelled data in order to fully exploit current and next-generation surveys.

Previous approaches in morphology classification have largely used fixed datasets of labelled galaxies acquired prior to model training.
This is true both for authors applying direct training \citep{Huertas-Company2015a, Sanchez2017, Fischer2018, Walmsley2018, HuertasCompany2018} and those applying transfer learning \citep{Ackermann2018, Perez-Carrasco2018, DominguezSanchez2019}.
Instead, we ask: to train the best model, which galaxies should volunteers label?

Selecting the most informative data to label is known as active learning. 
Active learning is useful when acquiring labels is difficult (expensive, time-consuming, requiring experts, private, etc).
This scenario is common for many, if not most, real-world problems. 
Terrestrial examples include detecting cardiac arrhythmia \citep{Rahhal2016}, sentiment analysis of online reviews \citep{Zhou2013}, and Earth observation \citep{Tuia2011, Liu2017}.
Astrophysical examples include stellar spectral analysis \citep{Solorio2005}, variable star classification \citep{Richards2012}, telescope design and time allocation \citep{Xia2016}, redshift estimation \citep{Hoyle2016} and spectroscopic follow-up of supernovae \citep{Ishida2018}.

\subsection{Active Learning Approach for Galaxy Zoo}

Given that only a small subset of galaxies can be labelled by humans, we should intelligently select which galaxies to label.
The aim is to make CNNs which are just as accurate without having to label as many galaxies.

Our approach is as follows. 
First, we train our CNN on a small randomly chosen initial training set.
Then, we repeat the following active learning loop:
\begin{enumerate}
    \item Measure the CNN prediction uncertainty on all currently-unlabelled galaxies (excluding a fixed test set)
    \item Apply an acquisition function (Section \ref{bald_and_mutual_information}) to select the most uncertain galaxies for labelling
    \item Upload these galaxies to Galaxy Zoo and collect volunteer classifications (in this work, simulated with historical classifications)
    \item Re-train the CNN and repeat
\end{enumerate}

Other astrophysics research has combined crowdsourcing with machine learning models.
\cite{Wright2017} classified supernovae in PanSTARRS \citep{Kaiser2010} by aggregating crowdsourced classifications with the predictions of expert-trained CNN and show that the combined human/machine ensemble outperforms either alone.
However, this approach is not directly feasible for Galaxy Zoo, where scale prevents us from recording crowdsourced classifications for every image.

A previous effort to consider optimizing task assignment was made by \cite{Beck2018}, who developed a `decision engine' to allocate galaxies for classification by either human or machine (via a random forest). 
Their system assigns each galaxy to the categories `Smooth' or `Featured'
\footnote{The actual categories used were `Featured' or `Not Featured' (Smooth + Artifact), but they argue that Artifact is sufficiently rare to not affect the results.}
, using SWAP \citep{Marshall2016} to decide how may responses to collect. 
This is in contrast to the system presented here which only requests responses for informative galaxies, but (for simplicity) requests the same number of responses for each informative galaxy.  
Another important difference is that \cite{Beck2018} train their model exclusively on galaxies which can be confidently assigned to a class,
 while the use of uncertainty in our model allows learning to occur from every classified galaxy.

This work is the first time active learning has been used for morphological classification, and the first time in astrophysics that active learning has been combined with CNNs or crowdsourcing.

In the following sections (\ref{bald_and_mutual_information}, \ref{estimating_mutual_information}, \ref{entropy_evaluation}), we derive an acquisition function that selects the most informative galaxies for labelling by volunteers.
We do this by combining the general acquisition strategy BALD \citep{MacKay1998, Houlsby2011} with our probabilistic model and Monte-Carlo Dropout \citep{Gal2016Uncertainty}.
We then use historical data to simulate applying active learning strategy to Galaxy Zoo (Section \ref{active_learning_application}) and compare the performance of models trained on galaxies selected using the mutual information versus galaxies selected randomly (Section \ref{active_learning_results}).

\subsection{BALD and Mutual Information}
\label{bald_and_mutual_information}

Bayesian Active Learning by Disagreement, BALD \citep{MacKay1998, Houlsby2011}, is a general information-theoretic acquisition strategy.
BALD selects subjects to label by maximising the mutual information between the model parameters $\theta$ and the probabilistic label prediction $y$.
We begin deriving our acquisition function by describing BALD and the mutual information.

We have observed data $\mathcal{D} = { ( x_{i}, y_i)}^n_{i=1}$. Here, $x_{i}$ is the $i$th subject and $y_i$ is the label of interest. 
We assume there are (unknown) parameters $\theta$ that model the relationship between input subjects $x$ and output labels $y$, $p(y|x, \theta)$. 
We would like to infer the posterior of $\theta$, $p(\theta | \mathcal{D})$. 
Once we know $p(y|x, \theta)$, we can make predictions on new galaxy images.

The mutual information measures how much information some random variable A carries about another random variable B, defined as:
\begin{equation}
    \mathbb{I}[A, B] = H[p(A)] - E_{p(B)}H[p(A|B)]
\end{equation}
where $H$ is the entropy operator and $E_{p(B)}H[p(A|B)]$ is the expected entropy of $p(A|B)$, marginalised over $p(B)$ \citep{Murphy2012}

We would like to know how much information each label $y$ provides about the model parameters $\theta$. 
We can then pick subjects $x$ to maximise the mutual information $\mathbb{I}[y, \theta]$, helping us to learn $\theta$ efficiently.
Substituting $A$ and $B$ for $x$ and $y$:

\begin{equation}
    \label{mutual_information}
    \mathbb{I}[y, \theta] = H[p(y|x,\mathcal{D})] - \mathbb{E}_{p(\theta|\mathcal{D})} [ H[p(y|x, \theta)] ]
\end{equation}

The first term is the entropy of our prediction for $x$ given the training data, implicitly marginalising over the possible model parameters $\theta$. 
We refer to this as the predictive entropy.
The predictive entropy reflects our overall uncertainty in $y$ given the training data available.

The second term is the expected entropy of our prediction made with a given $\theta$, sampling over each $\theta$ we might have inferred from $\mathcal{D}$.
The expected entropy reflects the typical uncertainty of each particular model on $x$.
Expected entropy has a lower bound set by the inherent difficulty in predicting $y$ from $x$, regardless of the available labelled data.

Confident disagreement between possible models leads to high mutual information.
For high mutual information, we should be highly uncertain about $y$ after marginalising over all the models we might infer (high $H[p(y|x,\mathcal{D})]$), but have each particular model be confident (low expected $H[p(y|x, \theta)]$ ).
If we are uncertain overall, but each particular model is certain, then the models must confidently disagree. 

Throughout this work, when we refer to galaxies as informative, we mean specifically that they have a high mutual information; they are \textit{informative for the model}.
These are not necessarily the galaxies which are the most \textit{informative for science}; any overlap will depend upon the research question at hand.
The scientific benefit of our approach is that we improve our morphological predictions for all galaxies using minimal newly-labelled examples.

\subsection{Estimating Mutual Information}
\label{estimating_mutual_information}

Rewriting the mutual information explicitly, replacing $y$ with our labels $k$ and $\theta$ with the network weights $w$:
\begin{equation}
    \label{explicit_mutual_information}
    \mathbb{I}[k, w] = \mathbb{H}[\int p(k|x, w) p (w | \mathcal{D}) dw] - \int p(w|\mathcal{D}) \mathbb{H}[p(k|x, w)] dw
\end{equation}

\cite{Gal2017} showed that we can use Eqn. \ref{dropout_approximation} to replace $p(w|\mathcal{D})$ in the mutual information (Eqn. \ref{explicit_mutual_information}):
\begin{equation}
    \label{mutual_information_with_dropout}
    \mathbb{I}[k, w] = \mathbb{H}[\int p(k|x, w) q^\ast dw] - \int q^\ast \mathbb{H}[p(k|x, w)] dw
\end{equation}

and again sample from $q^{\ast}$ with $T$ forward passes using dropout at test time (i.e. Monte Carlo integration):
\begin{equation}
    \mathbb{I}[k, w] = \mathbb{H}[\frac{1}{T} \sum_{t} p(k|x, w)] - \frac{1}{T} \sum_{t} \mathbb{H}[p(k|x, w)]
\end{equation}

Next, we need a probabilistic prediction for $k$,  $p(k|x, w)$. 
Here, we diverge from previous work.

Recall that we trained our network to make probabilistic predictions for $k$ by estimating the latent parameter $\rho$ from which $k$ is Binomially drawn (Eqn. \ref{model_likelihood}).
Substituting the probabilistic predictions of Eqn. \ref{model_likelihood} into the mutual information:
\begin{equation}
    \mathbb{I}[k, w] = \mathbb{H}[\frac{1}{T} \sum_{t} \text{Bin}(k|f^w(x), N)] - \frac{1}{T} \sum_{t} \mathbb{H}[\text{Bin}(k|f^w(x), N)]
\end{equation}
Or concisely:
\begin{equation}
    \label{concise_mutual_information}
    \mathbb{I}[k, w] = \mathbb{H}[\langle\text{Bin}(k|f^w(x), N)\rangle] - \langle \mathbb{H}[\text{Bin}(k|f^w(x), N)]\rangle
\end{equation}

A novel complication is that we do not know $N$, the total number of responses, prior to labelling. 
In GZ2, each subject is shown to a fixed number of volunteers, but (due to the decision tree) $N$ for each question will depend on responses to the previous question.
Further, technical limitations mean that even for the first question (`Smooth or Featured'), $N$ can vary (Figure \ref{vote_histograms}).
We (implicitly, for clarity) approximate $N$ with the expected $\langle N \rangle$ for that question.
In effect, we calculate our acquisition function with $N$ set to the value that, \textit{were we to ask volunteers to label this galaxy, we would expect} $N$ \textit{responses}. 

To summarise, Eqn. \ref{concise_mutual_information} asks: how much additional information would be gained about network parameters that we use to predict $\rho$ and $k$, were we to ask $\langle N \rangle$ people about subject $x$? 

\subsection{Entropy Evaluation}
\label{entropy_evaluation}

Having approximated $p(w|\mathcal{D})$ with dropout and calculated $p(k|x, w)$ with our probabilistic model, all that remains is to calculate the entropies $\mathbb{H}$ of each term.

$k$ is discrete and hence we can directly calculate the entropy over each possible state:
\begin{equation}
    \mathbb{H}[\text{Bin}(k|f^w(x), N)] = - \sum_{k=0}^{N} \text{Bin}(k|f^w(x), N) \log[ \text{Bin}(k|f^w(x), N)]
\end{equation}

For $\mathbb{H}[\langle\text{Bin}(k|f^w(x), N)\rangle]$, we can also enumerate over each possible $k$, where the probability of each $k$ is the mean of the posterior predictions (sampled with dropout) for that $k$:
\begin{equation}
    \begin{split}
    \mathbb{H}[\langle\text{B}&\text{in}(k|f^w(x), N)\rangle] = \\
    & - \sum_{k=0}^{N} \langle\text{Bin}(k|f^w(x), N)\rangle \log[ \langle\text{Bin}(k|f^w(x), N)\rangle]
    \end{split}
\end{equation}

and hence our final expression for the mutual information is:
\begin{equation}
    \begin{split}
        \mathbb{I}[k, w] & = \\
    & - \sum_{k=0}^{N} \langle\text{Bin}(k|f^w(x), N)\rangle \log[ \langle\text{Bin}(k|f^w(x), N)\rangle] \\
    & + \sum_{k=0}^{N} \text{Bin}(k|f^w(x), N) \log[ \text{Bin}(k|f^w(x), N)]
    \end{split}
\end{equation}

\subsection{Application}
\label{active_learning_application}

To evaluate our active learning approach, we simulate applying active learning during GZ2.
We compare the performance of our models when trained on galaxies selected using the mutual information versus galaxies selected randomly.
For simplicity, each simulation trains a model to predict either `Smooth or Featured' responses or `Bar' responses.

For the `Smooth or Featured' simulation, we begin with a small initial training set of 256 random galaxies.
We train a model and predict $p(k|\rho, N)$ (where $N$ is the expected number of volunteers to answer the question, calculated as the mean total number of responses for that question over all previous galaxies - see Figure \ref{vote_histograms}).
We then use our BALD acquisition function (Eqn. \ref{concise_mutual_information}) to identify the 128 most informative galaxies to label.
To simulate uploading the informative galaxies to GZ and receiving classifications, we retrieve previously collected GZ2 classifications.
Finally, we add the newly-labelled informative galaxies to our training set.
We refer to each execution of this process (training our model, selecting new galaxies to label, and adding them to the training set) as an \textit{iteration}.
We repeat for 20 iterations, recording the performance of our model throughout.

We selected 256 initial galaxies and 128 further galaxies per iteration, to match the training data size over which our `Smooth or Featured' model performance varies.
Our relatively shallow model reaches peak performance on around 3000 random galaxies; more galaxies do not significantly improve performance.

For the `Bar' simulation, we observe that performance saturates after more galaxies (approx. 6000) and so we double the scale; we start with 512 galaxies and acquire 256 further galaxies per iteration.
This matches previous results (and intuition) that `Smooth or Featured' is an easier question to answer than `Bar'. 
Identifying bars, particularly weak bars, is challenging for both humans \citep{Masters2012, Kruk2018} and machines (including CNNs, \citealt{Sanchez2017}).

To measure the effect of our active learning strategy, we also train a baseline classifier by providing batches of randomly selected galaxies. 
We aim to compare two acquisition strategies for deciding which galaxies to label: selecting galaxies with maximal mutual information (active learning via BALD and MC Dropout) or selecting randomly (baseline).
We evaluate performance on a fixed test set of 2500 random galaxies.
We repeat each simulation four times to reduce the risk of spurious results from random variations in performance. 

\subsection{Results}
\label{active_learning_results}

\begin{figure}
    \textbf{GZ2 `Smooth' Active Learning Performance}
    \centering
    \includegraphics[width=0.4\textwidth]{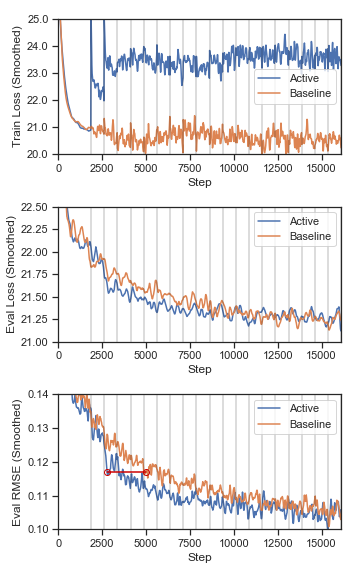}
    \caption{
        Training loss (upper), evaluation loss (middle), and RMSE (lower) of model performance on `Smooth or Featured' during active learning simulations, by iteration (set of new galaxies).
        Vertical bars denote new iterations, where new galaxies are acquired and added to the training set. 
        Prior to 2000 training iterations, both the random selection (baseline) models and active learning models train on only the initial random training set of 256 galaxies, and hence show similar performance.
        Around 2000 to 3500 iterations, after acquiring 128-256 additional galaxies, the active learning model shows a clear improvement in evaluation performance over the baseline model.
        We annotate in red where each model achieves the maximal relative RMSE improvement, highlighting the reduction in newly-labelled galaxies required (vertical bars = 128 new galaxies).
        Note that active learning leads to a dramatically higher training loss, indicating that more challenging galaxies are being identified as informative and added to the training set.
        }
    \label{smooth_active_performance}
\end{figure}

\begin{figure}
    \textbf{GZ2 `Bar' Active Learning Performance}
    \centering
    \includegraphics[width=0.4\textwidth]{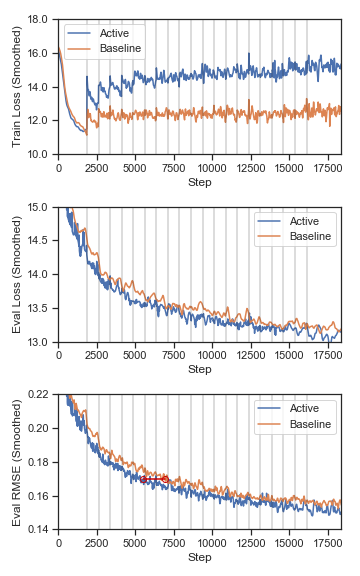}
    \caption{
        As with \ref{smooth_active_performance}, but for the `Bar' active learning simulations.
        Again, active learning leads to a clear improvement in evaluation performance and a dramatically higher training loss (indicating challenging galaxies are being selected).
        We annotate in red where each model achieves the maximal relative RMSE improvement, highlighting the reduction in newly-labelled galaxies required (vertical bars = 256 new galaxies).
    }
    \label{bar_active_performance}
\end{figure}

For both `Smooth' and `Bar' simulations, our probabilistic models achieve equal performance on fewer galaxies using active learning versus random galaxy selection.
We show model performance by iteration for the `Smooth' (Figure \ref{smooth_active_performance}) and `Bar' (Figure \ref{bar_active_performance}) simulations.
We display three metrics: training loss (model surprise on previously-seen images, measured by Eqn. \ref{binomial_loss_vf}), evaluation loss (model surprise on unseen images), and root-mean-square error (RMSE).
We measure the RMSE between our maximum-likelihood-estimates $\hat{\rho}$ and $\rho_{\text{proxy}} = \frac{k}{N}$ as $\rho$ itself is never observed and hence cannot be used for evaluation.
Due to the high variance in metrics between batches, we smooth our metrics via LOWESS \citep{Cleveland1979} and average across 4 simulation runs. 

For `Smooth', we achieve equal RMSE scores with, at best, $\sim 60$\% fewer newly-labelled galaxies (RMSE of 0.117 with 256 vs. 640 new galaxies, Figure \ref{smooth_active_performance}).
Similarly for `Bar', we achieve equal RMSE scores with, at best, $\sim 35$\% fewer newly-labelled galaxies (RMSE of 0.17 with 1280 vs. 2048 new galaxies, Figure \ref{bar_active_performance}).
Active learning outperforms random selection in every run.

Given sufficient ($\sim 3000$ for `Smooth', $\sim 6000$ for `Bar') galaxies, our models eventually converge to similar performance levels -- regardless of galaxy selection.
We speculate that this is because our relatively shallow model architecture places an upper limit on performance.
In general, model complexity should be large enough to exploit the information in the training set yet small enough to avoid fitting to spurious patterns.
Model complexity increases with the number of free parameters, and decreases with regularization \citep{Friedman2001}.
Our model is both shallow and well-regularized (recall that dropout was originaly used as a regularization technique, Section \ref{monte_carlo_dropout}).
A more complex (deeper) model may be able to perform better by learning from additional galaxies.

\subsubsection{Selected Galaxies}

Which galaxies do the models identify as informative?
To investigate, we randomly select one `Smooth or Featured' and one `Bar' simulation.

\begin{figure}
    \centering
    \includegraphics[width=0.66\columnwidth, trim={0 0 0 0.5cm}]{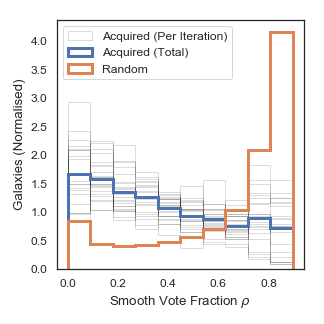}
    \includegraphics[width=0.8\columnwidth,trim={0 1cm 12cm 2cm},clip]{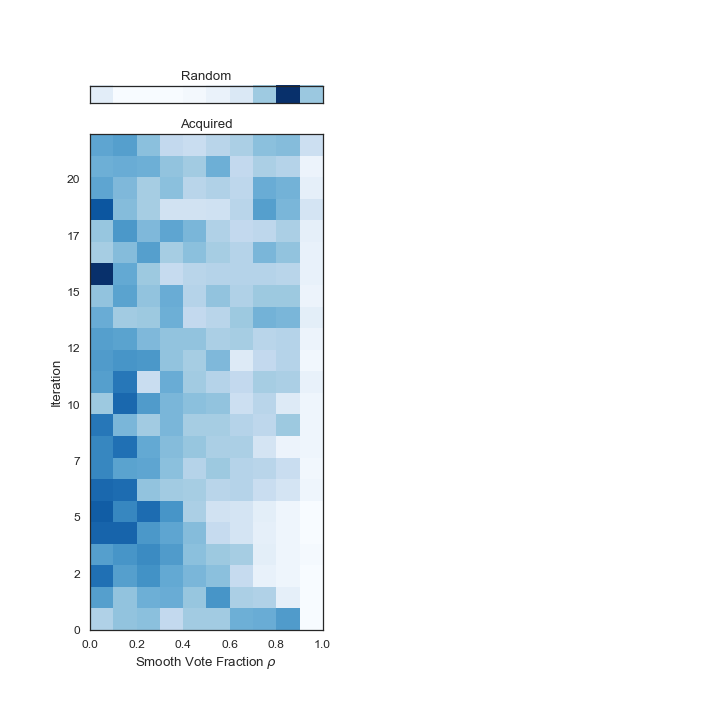}
    \caption{
        Distribution of observed `Smooth' vote fraction $p$ in galaxies acquired during Galaxy Zoo `Smooth or Featured' active learning simulation.
        Above: Distribution of acquired $p$ over all iterations, compared against random selection. 
        While randomly selected galaxies are highly smooth, our acquisition function selects galaxies from across the $p$ range, with a moderate preference towards featured.
        Below: Distribution of $p$ by iteration, compared against random selection (upper inset). 
        Our acquisition function strongly prefers featured galaxies in early ($n < \sim 7$) iterations, and then selects a more balanced sample.
        This likely compensates for the initial training sample being highly smooth.
    }
    \label{smooth_selected_galaxies}
\end{figure}

For the `Smooth or Featured' simulation, Figure \ref{smooth_selected_galaxies} shows the observed `Smooth' vote fraction distribution, per iteration (set of new galaxies) and in total (summed over all new galaxies).
Highly smooth galaxies are common in the general GZ2 catalogue. 
Random selection therefore leads to a training sample skewed towards highly smooth galaxies.
In contrast, our acquisition function is far more likely to select galaxies which are featured, leading to a more balanced sample.
This is especially true for the first few iterations; we speculate that this counteracts the skew towards smooth in the randomly selected initial training sample.
By the final training sample, featured galaxies become moderately more common than smooth (mean $\frac{k_{\text{smooth}}}{N}$ = 0.38).
This suggests that featured galaxies are (on average) more informative for the model -- over and above correcting for the skewed initial training sample.
We speculate that featured galaxies may be more visually diverse, leading to a greater challenge in fitting volunteer responses, more disagreement between dropout-approximated-models, and ultimately higher mutual information.

\begin{figure}
    \centering
    \includegraphics[width=0.66\columnwidth,trim={0 0 0 0.5cm},clip]{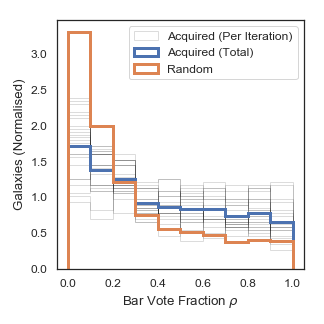}
    \includegraphics[width=0.8\columnwidth,trim={0 1cm 12cm 2cm},clip]{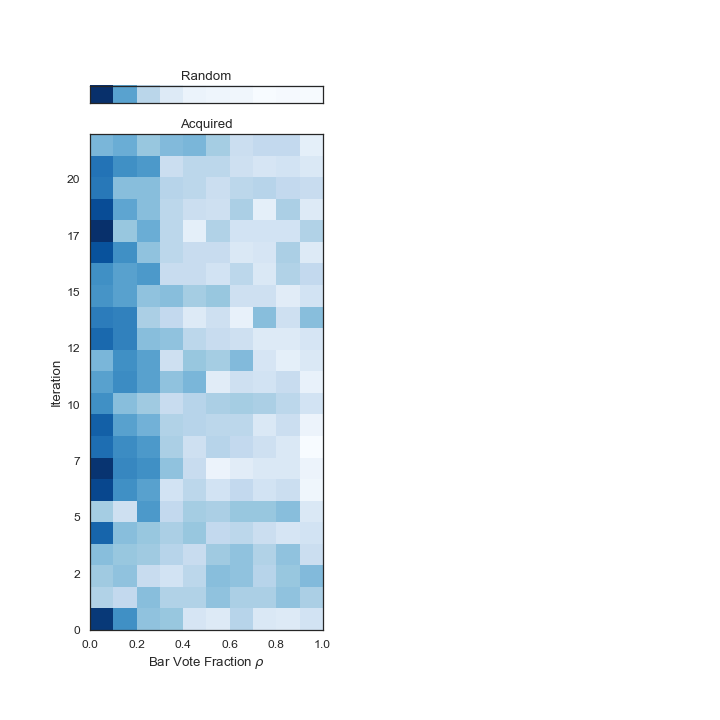}
    \includegraphics[width=0.66\columnwidth,trim={0 0 0 0.5cm},clip]{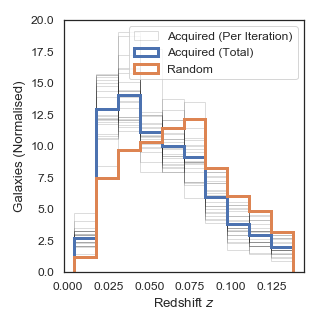}
    \caption{
        Upper: Distribution of observed `Bar' vote fraction $p$ in galaxies acquired during Galaxy Zoo `Bar' active learning simulation.
        While randomly selected galaxies are highly non-barred, the `Bar' model selects a more balanced sample.
        Middle: Distribution of `Bar' $p$ by iteration, compared against random selection (upper inset). 
        Our acquisition function selects a similar $rho$ distribution at each iteration.
        Lower: Redshift distribution of acquired galaxies over all iterations, compared against random selection. 
        The `Bar' model selects lower redshift galaxies, which are both more featured and better resolved (i.e. less visually ambiguous).
    }
    \label{bar_selected_galaxies}
\end{figure}

For the `Bar' simulation, Figure \ref{bar_selected_galaxies} shows the `Bar' vote fraction distribution, per iteration and in total, as well as the total redshift distribution.
Again, our acquisition function selects a more balanced sample by prioritising (rarer) barred galaxies.
This selection remains approximately constant (within statistical noise) as more galaxies are acquired.
With respect to redshift, our acquisition function prefers to select galaxies at lower redshifts. 
Based on inspection of the selected images (Figure \ref{informative_bar}), we suggest that these galaxies are more informative to our model because such galaxies are better resolved (i.e. less ambiguous) and more likely to be barred.

\begin{figure*}
    \centering
    \begin{subfigure}[b]{\columnwidth}
        \includegraphics[width=0.9\columnwidth]{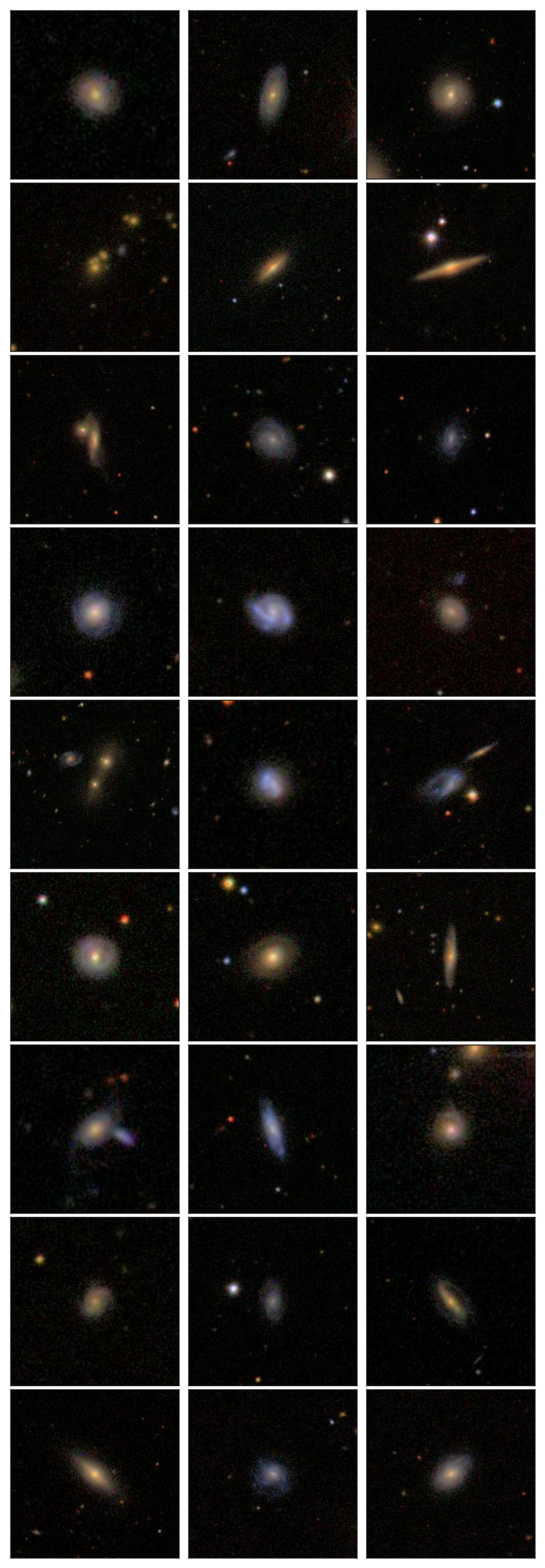}
        \caption{Galaxies with maximum mutual information for `Smooth or Featured'}
    \end{subfigure}
    \begin{subfigure}[b]{\columnwidth}
        \includegraphics[width=0.9\columnwidth]{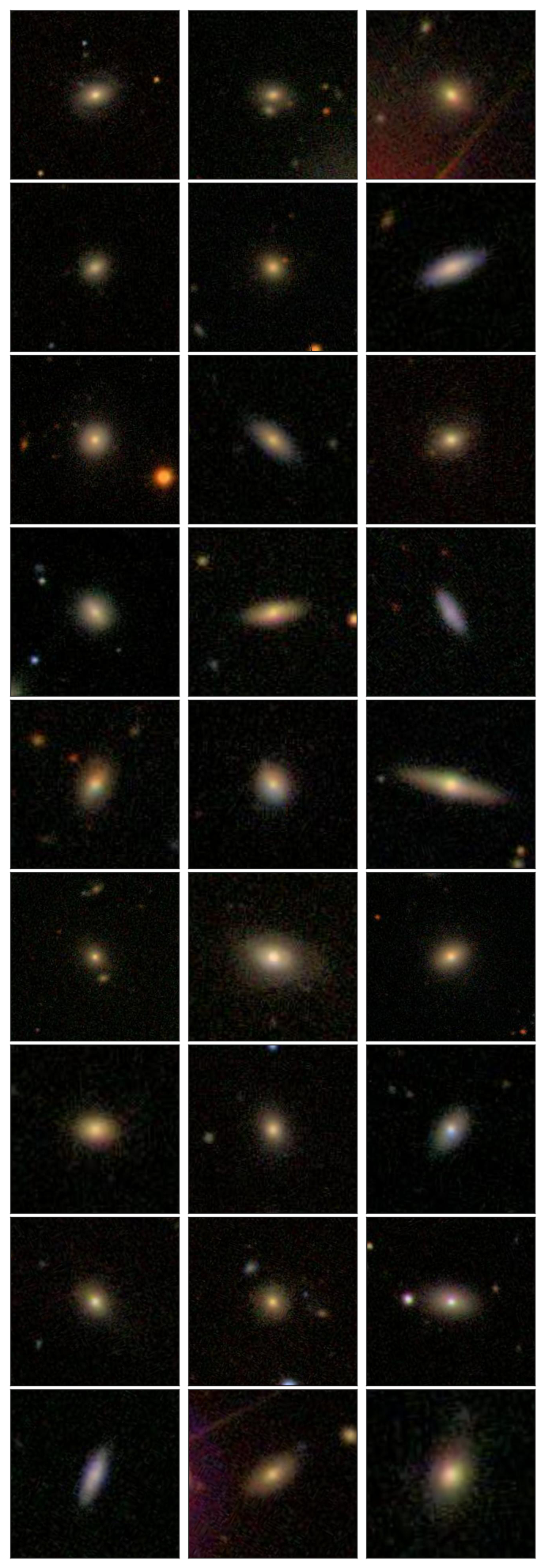}
        \caption{Galaxies with minimum mutual information for `Smooth or Featured'}
    \end{subfigure}
    \caption{
        Informative and uninformative galaxies from the (hidden) test subset, as identified by our novel acquisition function and the final model from a `Smooth or Featured' simulation. 
        When active learning is applied to Galaxy Zoo, volunteers will be more frequently presented with the most informative images (left) than the least (right).
        }
    \label{informative_smooth}
\end{figure*}

\begin{figure*}
    \centering
    \begin{subfigure}[b]{\columnwidth}
        \includegraphics[width=0.9\columnwidth]{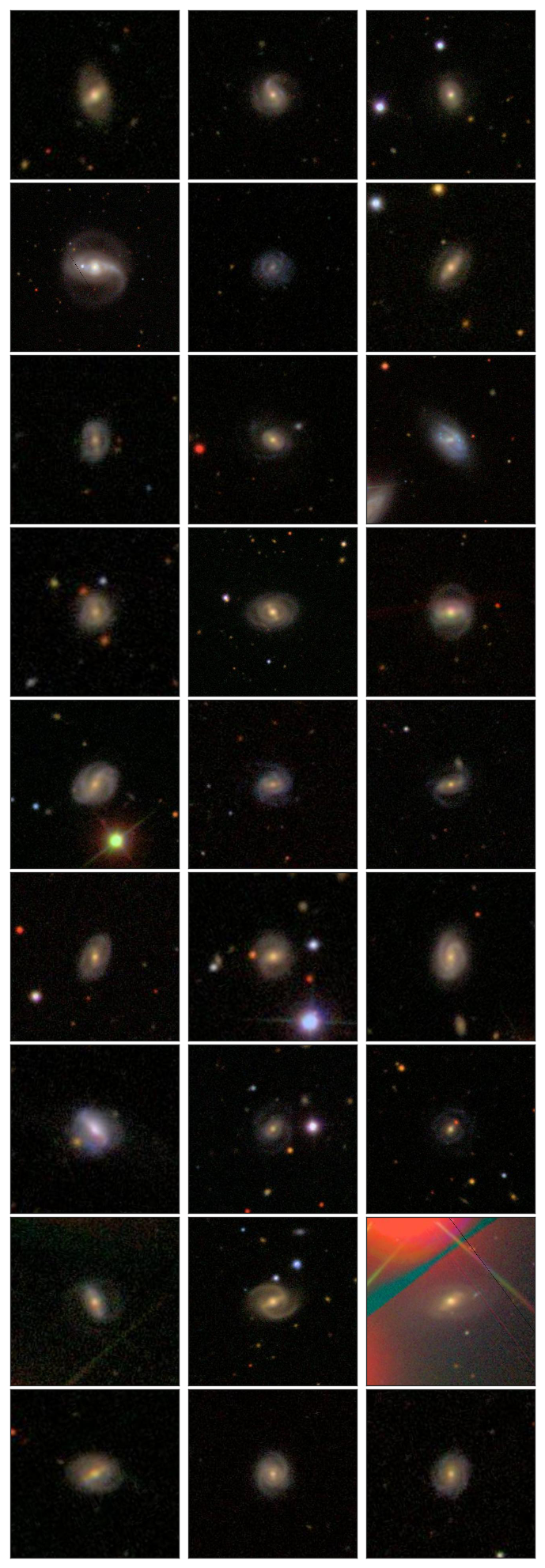}
        \caption{Galaxies with maximum mutual information for `Bar'}
    \end{subfigure}
    \begin{subfigure}[b]{\columnwidth}
        \includegraphics[width=0.9\columnwidth]{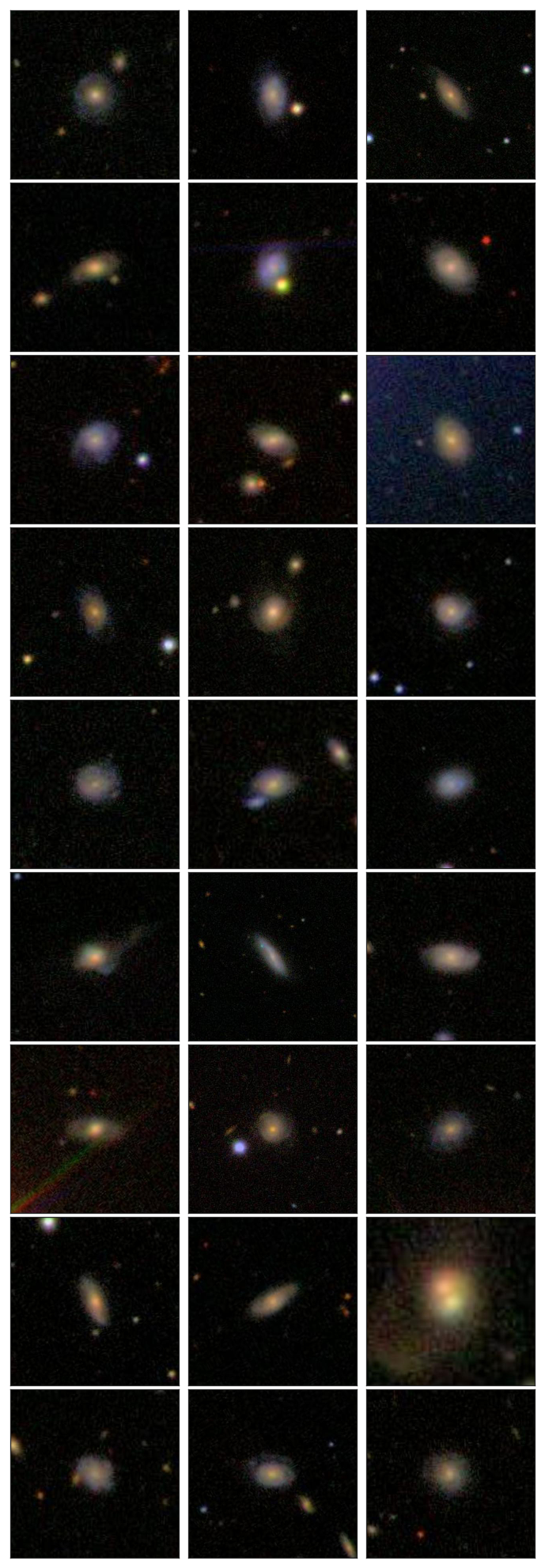}
        \caption{Galaxies with minimum mutual information for `Bar'}
    \end{subfigure}
    \caption{
        As with Figure \ref{informative_smooth} above, but showing galaxies identified by the final model from a `Bar' simulation.
    }
    \label{informative_bar}
\end{figure*}

We present the most and least informative galaxies from the (fixed and never labelled) test subset for `Smooth' (Figure \ref{informative_smooth} and Bar (Figure \ref{informative_bar}), as identified by our novel acquisition function and the final models from each simulation.\\

\section{Discussion}
\label{discussion}

Learning from fewer examples is an expected benefit of both probabilistic predictions and active learning.
Our models approach peak performance on remarkably few examples: 2816 galaxies for `Smooth' and 5632 for `Bar'.
With our system, volunteers could complete Galaxy Zoo 2 in weeks\footnote{For example, classifying $\sim$ 10,000 galaxies (sufficient to train our models to peak performance) at the mean GZ2 classification rate of $\sim$ 800 galaxies/day would take $\sim$ 13 days.} rather than years if the peak performance of our models would be sufficient for their research.
Further, reaching peak performance on relatively few examples indicates that an expanded model with additional free parameters is likely to perform better \citep{Murphy2012}.

For this work, we rely on GZ2 data where $N$ (the number of responses to a galaxy) is unknown before making a (historical) classification request.
Therefore, when deriving our acquisition function, we approximated $N$ as $\langle N\rangle$ (the expected number of responses).
However, during live application of our system, we can control the Galaxy Zoo classification logic to collect exactly $N$ responses per image, for any desired $N$.
This would allow our model to request (for example) one more classification for \textit{this} galaxy, and three more for \textit{that} galaxy, before retraining.
Precise classification requests from our model will enable us to ask volunteers exactly the right questions, helping them make an even greater contribution to scientific research.

We also hope that this human-machine collaboration will provide a better experience for volunteers.
Inspection of informative galaxies (Figures \ref{smooth_selected_galaxies}, \ref{bar_selected_galaxies}) suggests that more informative galaxies are more diverse than less informative galaxies.
We hope that volunteers will find these (now more frequent) informative galaxies interesting and engaging.

Our results motivate various improvements to the probabilistic morphology models we introduce.
In Section \ref{probabilistic_results}, we showed that our models were approximately well-calibrated, particularly after applying MC Dropout. 
However, the calibration was imperfect; even after applying MC Dropout, our models remain slightly overconfident (Figure \ref{coverage_comparison}). 
We suggest two reasons for this remaining overconfidence.
First, within the MC Dropout approximation, the dropout rate is known to affect the calibration of the final model \citep{Gal2017a}.
We choose our dropout rate arbitrarily ($0.5$); however, this rate may not sufficiently vary the model to approximate training many models.
One solution is to `tune' the dropout rate until the calibration is correct \citep{Gal2017a}.
Second, the MC Dropout approximation is itself imperfect; removing random neurons with dropout is not identical to training many networks.
As an alternative, one could simply train several models and ensemble the predictions \citep{Lakshminarayanan2016}.
Both of these approaches are straightforward given a sufficient computational budget.

We also showed the distribution of model predictions over all galaxies generally agrees well with the distribution of predictions from volunteers (i.e. we are globally unbiased, Section \ref{probabilistic_results}). 
However, we noted that the models are `reluctant' to predict extreme $\rho$  (the typical response probability, Section \ref{probabilistic_framework}).
We suggest that this is a limitation of our generative model for volunteer responses.
The binomial likelihood becomes narrow when $p$ (here, $\rho$) is extreme, and hence the model is heavily penalised for incorrect extreme $p$ estimates.
If volunteer responses were precisely binomially distributed (i.e. $N$ independent identically-distributed trials per galaxy, each with a fixed $p$ of a positive response), this heavy penalty would correctly reflect the significance of the error.
However, our binomial model of volunteers is only approximate; one volunteer may give consistently different responses to another.
In consequence, the true likelihood of non-extreme $k$ responses given $\rho$ is wider than the binomial likelihood from the `typical' response probability $\rho$ suggests, and the network is penalised `unfairly'.
The network therefore learns to avoid making risky extreme predictions. 

If this suggestion is correct, the risk-averse prediction shift will be monotonic (i.e. extreme galaxies will have slightly different $\rho$ but still be ranked in the same order) and hence researchers selecting galaxies near extreme $\rho$ may simply choose a slightly higher or lower $\hat{\rho}$ threshold.
To resolve this issue, one could apply a monotonic rescaling to the network predictions (as we do in Appendix A), introduce a more sophisticated model of volunteer behaviour \citep{Marshall2016,Beck2018,Dickinson2019}, or calibrate the loss to reflect the scientific utility of extreme predictions \citep{Cobb2018}.
As predictions are globally unbiased for all non-extreme $\rho$, and extreme $\rho$ predictions can be corrected post hoc (above), our network is ready for use.

Throughout this work, our goal has been to predict volunteer responses at scale.
These responses are known to vary systematically with e.g. redshift \citep{Willett2013,Hart2016} and colour \citep{Cabrera_Vives_2018}, and hence require calibration prior to scientific analysis.
Unlike \cite{Sanchez2017} and \cite{Khan2018}, who train on redshift-calibrated `debiased' responses, we expect and intend to reproduce these systematics. 
We prefer to apply calibration methods to our predictions.
A calibrated CNN-predicted catalogue will be presented as part of a future Galaxy Zoo data release.

Finally, we highlight that our approach is highly general.
We hope that Bayesian CNNs and active learning can contribute to the wide range of astrophysical problems where CNNs are applicable (e.g. images, time series), uncertainty is important, and the data is expensive to label, noisy, imbalanced, or includes rare objects of interest.
In particular, imbalanced datasets (where some labels are far more common than others) are common throughout astrophysics.
Topics include transient classification \citep{Wright2017}, fast radio burst searches \citep{Zhang2018}, and exoplanet detection \citep{Osborn2019}.
Active learning is known to be effective at correcting such imbalances \citep{Ishida2018}.
Our results suggest that this remains true when active learning is combined with CNNs (this work is the first astrophysics application of such a combination).
Recall that smooth galaxies are far more common in GZ2 but featured galaxies are strongly preferentially selected by active learning -- automatically, without our instruction -- apparently to compensate for the imbalanced data (Figure \ref{smooth_selected_galaxies}).
If this observation proves to be general, we suggest that Bayesian CNNs and active learning can drive intelligent data collection to overcome research challenges throughout astrophysics.

\section{Conclusion}
\label{conclusion}

Previous work on predicting visual galaxy morphology with deep learning has either taken no account of uncertainty or trained only on confidently-labelled galaxies.
Our Bayesian CNNs model and exploit the uncertainty in Galaxy Zoo volunteer responses using a novel generative model of volunteers.
This enables us to accurately answer detailed morphology questions using only sparse labels ($\sim 10$ responses per galaxy).
Our CNNs can also express uncertainty by predicting probability distribution parameters and using Monte-Carlo Dropout \citep{Gal2017}.
This allows us to predict posteriors for the expected volunteer responses to each galaxy.
These posteriors are reliable (i.e. well-calibrated), show minimal systematic bias, and match or outperform previous work when reduced to point estimates (for comparison).
Using our posteriors, researchers will be able to draw statistically powerful conclusions about the relationships between morphology and AGN, mass assembly, quenching, and other topics.

Previous work has also treated labelled galaxies as a fixed dataset from which to learn.
Instead, we ask: which galaxies should we label to train the best model?
We apply active learning \citep{Houlsby2011} -- our model iteratively requests new galaxies for human labelling and then retrains. 
To select the most informative galaxies for labelling, we derive a custom acquisition function for Galaxy Zoo based on BALD \citep{MacKay1998}. 
This derivation is only possible using our posteriors.
We find that active learning provides a clear improvement in performance over random selection of galaxies.
The galaxies identified as informative are generally more featured (for the `Smooth or Featured' question) and better resolved (for the `Bar' question), matching our intuition.

As modern surveys continue to outpace traditional citizen science, probabilistic predictions and active learning become particularly crucial.
The methods we introduce here will allow Galaxy Zoo to produce visual morphology measurements for surveys of any conceivable scale on a timescale of weeks.
We aim to launch our active learning strategy on Galaxy Zoo in 2019.

\section*{Acknowledgements}

MW would like to thank H. Dom\'inguez Sanchez and M. Huertas-Company for helpful discussions.

MW acknowledges funding from the Science and Technology Funding Council (STFC) Grant Code ST/R505006/1.
We also acknowledge support from STFC under grant ST/N003179/1. 
LF, CS, HD and DW acknowledge partial support from one or more of the US National Science Foundation grants IIS-1619177, OAC-1835530, and AST-1716602.

This research made use of the open-source Python scientific computing ecosystem, including SciPy \citep{scipy}, Matplotlib \citep{Hunter2007}, scikit-learn \citep{Pedregosa2011}, scikit-image \citep{VanderWalt2014} and Pandas \citep{McKinney2010}.

This research made use of Astropy, a community-developed core Python package for Astronomy \citep{TheAstropyCollaboration2013,TheAstropyCollaboration2018}.

This research made use of TensorFlow \citep{tensorflow2015-whitepaper}.

All code is publicly available on GitHub at \href{www.github.com/mwalmsley/galaxy-zoo-bayesian-cnn}{www.github.com/mwalmsley/galaxy-zoo-bayesian-cnn} \citep{Walmsley2019}.

\section*{Appendix A}
\appendix
\label{appendix}

\makeatletter 
\renewcommand{\thefigure}{A\@arabic\c@figure}
\makeatother

CNN predictions are not (in general) well-calibrated probabilities \citep{Lakshminarayanan2016, Guo2017}.
Interpreting them as such may cause systematic errors in later analysis.
To illustrate this problem, we show how the CNN probabilities published in DS+18 \citep{Sanchez2017} significantly overestimate the prevalance of expert-classified barred galaxies.
We chose DS+18 as the most recent deep learning morphology catalogue made publicly available, and thank the authors for their openness.
We do not believe this issue is unique to DS+18. 
We highlight this issue not as a criticism of DS+18 specifically, but to emphasise the advantages of using probabilistic methods.

DS+18 trained a CNN to predict the probability that a galaxy is barred (DS+18 Section 5.3).
Barred galaxies were defined as those galaxies labelled as having any kind of bar (weak/intermediate/strong) in expert catalogue \citealt{Nair2010} (N10).
We refer to such galaxies as Nair Bars.
We chose to investigate this particular DS+18 model because it explicitly aims to reproduce the (expert) N10 classifications, allowing for direct comparison of the predicted probabilities against the true labels. 

We first show that these CNN probabilities are not well-calibrated.
We then demonstrate a simple technique to infer probabilities for Nair Bars from GZ2 vote fractions.
Finally, we show that, as our Bayesian CNN estimates of GZ2 vote fractions are well-calibrated, these vote fractions can be used to estimate probabilities for Nair Bars.
The practical application is to predict what \cite{Nair2010} would have recorded, had the expert authors visually classified every SDSS galaxy.

\begin{figure}
    \centering
    \includegraphics[width=\columnwidth]{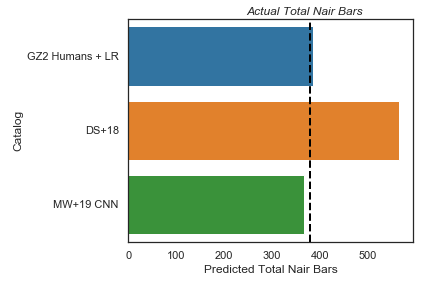}
    \caption{
        Predictions for the total number of galaxies labelled as `Bar' by human expert N10 in test galaxy subset (correct answer: 379).
        DS+18 overestimates the number of Nair Bars (559).
        We find that GZ2 vote fractions from volunteers can be used to make an improved estimate (396) with a rescaling correction calculated via logistic regression (GZ2 Humans + LR).
        Applying the same correction to the vote fractions predicted by the Bayesian CNN in this work (MW+19) also produces an improved estimate (372).
        By accurately predicting the vote fractions, and then applying a correction to map from vote fractions to expert responses, we can predict what N10 would have said for the full SDSS sample.
        }
    \label{total_bars}
\end{figure}

\begin{figure}
    \centering
    \includegraphics[width=\columnwidth]{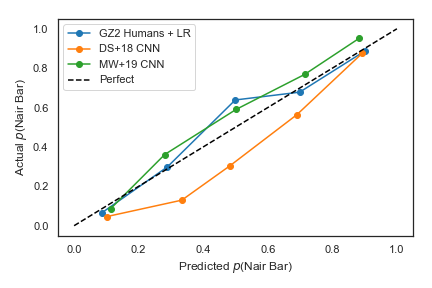}
    \includegraphics[width=\columnwidth]{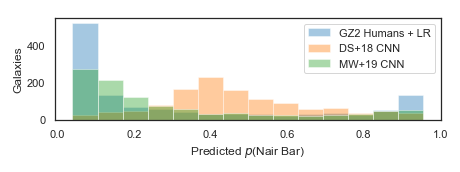}
    \caption{
        Above: Comparison of calibration curves for each predictive model. 
        The calibration curve is calculated by binning the predicted probabilities and counting the fraction of Nair Bars in each bin.
        The fraction of Nair Bars in a given bin approximates the true (frequentist) probability of each binned galaxy being a Nair Bar.
        Points compare the predicted fraction of Nair Bars ($x$ axis) with the actual fraction ($y$ axis) for 5 equally-spaced bins.
        For well-calibrated probabilities, the predicted and actual fractions are equal (black dashed line).
        Below: the distribution of Nair Bar predictions from each model. 
        DS+18 typically predicts $p \sim 0.4$ (below) and has a relatively poor calibration near $p \sim 0.4$ (above), leading to a significant overestimate of the total number of Nair Bars.
    }
    \label{bar_calibration_curves}
\end{figure}

We select a random subset of 1211 galaxies classified by N10 (this subset is motivated below). 
How many barred galaxies are in this subset?
The DS+18 Nair Bar `probabilities' $p_i$ (for each galaxy $i$) predict $\sum_i p_i = 559$ Nair Bars.
However, only 379 are actually Nair Bars (Figure \ref{total_bars}).
This error is caused by the DS+18 Nair Bar `probabilities' being, on average, skewed towards predicting `Bar', as shown by the calibration curve of the DS+18 Nair Bar probabilities (Figure \ref{bar_calibration_curves}).

\begin{figure}
    \centering
    \includegraphics[width=\columnwidth]{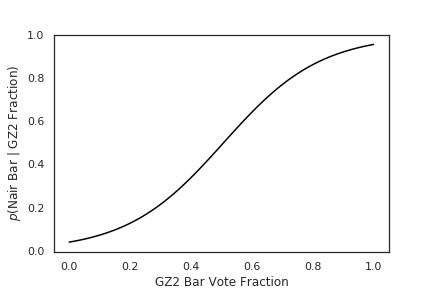}
    \caption{
        The rescaling function used to map GZ2 vote fractions to $p(\text{Nair Bar} | \text{GZ2 Fraction})$, estimated via logistic regression. 
        This rescaling function is also used (without modification) to map Bayesian CNN GZ2 vote fraction predictions to $p(\text{Nair Bar} | \text{BCNN-predicted GZ2 Fraction})$}
    \label{rescaling_function}
\end{figure}

How can we better predict the total number of Nair Bars?
GZ2 collected volunteer responses for many galaxies classified by N10 
(6,051 of 14,034 match within $5^{\prime\prime}$, after filtering for total `Bar?' votes $N_{\text{bar}} > 10$ as in Section \ref{experimental_setup}).
The fraction of volunteers who responded `Bar' to the question `Bar?' is predictive of Nair Bars, but is not a probability \citep{Lintott2008}. 
For example, volunteers are less able to recognise weak bars than experts \citep{Masters2012}, and hence the `Bar' vote fraction only slightly increases for galaxies with weak Nair Bars vs. galaxies without.
We need to rescale the GZ2 vote fractions.
To do this, we divide the N10 catalogue into 80\% train and 20\% test subsets and use the train subset to fit (via logistic regression) a rescaling function (Figure \ref{rescaling_function}) mapping GZ2 vote fractions to $p(\text{Nair Bar} | \text{GZ2 Fraction})$.
We then evaluate the calibration of these probabilities on the test subset, which is the subset of 1211 galaxies used above. 
We predict 396 Nair Bars, which compares well with the correct answer of 379 vs. the DS+18 answer of 559 (Figure \ref{total_bars}).
This directly demonstrates that our rescaled GZ2 predictions are correctly calibrated over the full test subset.
The calibration curve shows no systematic skew, unlike DS+18 (Figure \ref{bar_calibration_curves}).

Since the GZ2 vote fractions can be rescaled to Nair Bar probabilities, and the Bayesian CNN makes predictions of the GZ2 vote fractions, we can also rescale the Bayesian CNN predictions into Nair Bar probabilities using the same rescaling function. 
The rescaled Bayesian CNN GZ2 vote predictions correctly estimate the count of Nair Bars (372 bars predicted vs. 379 observed bars, Figure \ref{total_bars}).

\begin{figure}
    \centering
    \includegraphics[width=\columnwidth]{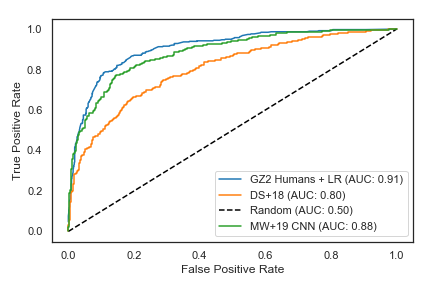}
    \caption{Comparison of ROC curves for predicting Nair Bars using each model.}
    \label{nair_bar_roc}
\end{figure}

Finally, we note that if the research goal is simply to identify samples of e.g. Nair Bars, one can do so by interpreting each prediction as a score (i.e. an arbitrary scalar, as opposed to a probability).
When interpreted as scores, the rescaled GZ2 votes - both observed from volunteers and predicted by the Bayesian CNN - outperform DS+18 in identifying Nair Bars at all thresholds (Figure \ref{nair_bar_roc}). 
This may be because our BCNN can learn to detect bars from the extensive GZ2 sample (56,048 galaxies with $N_{bar} \geq 10$) before those predictions are rescaled to correspond to Nair Bars, rather than DS+18's approach of training only on the much smaller set of galaxies (7000) directly labelled in \cite{Nair2010}.

Nair Bars are initially defined through repeated expert classification (as close to `gold standard' ground truth as exists for imaging data) and hence accurate automated identification of Nair Bars is directly useful for morphology research.

\section*{Appendix B - Theoretical Background on Variational Inference}
\appendix
\label{appendix_b}

The general problem of Bayesian inference can be framed in terms of a probabilistic model where we have some observed random variables $Z$ and some latent variables $\theta$ and we wish to infer $P(\theta \mid Z)$ after observing some data. Our probabilistic model $P(Z, \theta)$ allows us to use Bayes rule to do so; $P(\theta \mid Z) = \frac{P(\theta , Z)}{P(Z)} = \frac{P(Z \mid \theta)p(\theta)}{P(Z)}$.
In the setting of discriminative learning, the observed variables are the inputs and outputs of our classification task $X$ and $Y$, and we directly parameterise the distribution $P(y \mid x, \theta)$ in order to make predictions by marginalising over the unknown weights, that is, the prediction for an unseen point $x$ given training data $X$ is 

\begin{align}
p(y \mid x, X, Y) = \int p(y \mid x, \theta) p(\theta \mid X, Y) d\theta
\end{align}

While this is a simple framework, in practice the integrals required to normalise Bayes' rule and to take this marginal are often not analytically tractable, and we must resort to numerical approaches.

While there are many possible ways to perform approximate Bayesian inference, here we will focus on the framework of \textit{variational inference}.
The essential idea of variational inference is to approximate the posterior $P(\theta \mid Z)$ with a simpler distribution $q(\theta)$ which is `as close as possible' to $P(\theta \mid Z)$, and then use $q$ in place of the posterior.
This can take the form of analytically finding the optimal $q$ subject only to some factorisation assumptions using the tools of the calculus of variations, but the case that is relevant to our treatment is when we fix $q$ to be some family of distributions $q_\xi(\theta)$ parameterised by $\xi$ and fit $\xi$, changing an integration problem to an optimisation one.

The measure of `as close as possible' used in variational inference in the Kullback-Leibler (KL) divergence, or the relative entropy, a measure of distance between two probability distributions defined as 

\begin{align}
D_{KL}(p:q) = \int p(x) (\log p(x) - \log q(x)) dx
\end{align}

The objective of variational inference is to choose the $q$ such that $D_{KL}(q(\theta): p(\theta \mid X))$ is minimised.
Minimising this objective can be shown to be equivalent to maximising the 'log Evidence Lower BOund', or ELBO,

\begin{align}
        L(q) = \mathbb{E}_{q(\theta)} - [ \log p ( Y \mid X, \theta) p(\theta)  - \log q(\theta)]
\end{align}

The reason for the name is the relationship 

\begin{align}
\log P(X) = D_{KL}(q(\theta) : p(\theta \mid X)) + L(q)
\end{align}

which implies, since the KL divergence is strictly positive, that $L$ provides a lower bound on the log of the evidence $P(X)$, the denominator in Bayes rule above.
By optimising the parameters of $q$ $xi$, with respect to $L$, one can find the best approxmation to the posterior in the family of parameterised distributions chosen in terms of the ELBO.

The key advantage of this formalism is that the ELBO only involves the tractable terms of the model, $P(X \mid \theta)$ and $P(\theta)$. 
The expectation is over the approximating distribution, but since we are able to choose $q$ we can make a choice that is easy to sample from, and therefore it is straightforward to obtain a monte carlo approximation of $L$ via sampling, which is sufficient to obtain stochastic gradients of $L$ which can be used for optimisation.
The integral over the posterior on $\theta$ in the marginalisation step can likewise be approximated via sampling from $q$ if neccesary.

For neural networks, a common approximating distribution is dropout \citep{Srivastava2014}.
The dropout distribution over the weights of a single neural network layer is parameterised by a weight matrix $M$ and a dropout probability $p$.
Draws from this distribution are described by 

\begin{align}
  W_{ij} = M_{ij} z_{j}
\end{align}

where $z_j \sim \rm{Bernoulli}(p)$.
\cite{Gal2016Uncertainty} introduced approximating $p(w|\mathcal{D})$, with a dropout distributions over the weights of a network, and showed that in this case optimising the standard likelihood based loss is equivalent to the variational objective that would be obtained for the dropout distribution, so we may interpret the dropout distribution over the weights of a trained model as an approximation to the posterior distribution $p(w \mid \mathcal{D})$.

We can use this approximating distribution as a proxy for the true posterior when we marginalise over models to make predictions;
\begin{equation}
    \int p(k|x, w) p (w | \mathcal{D}) dw \approx \int p(k|x, w) q^\ast dw
\end{equation}

A more detailed mathematical exposition of dropout as variational inference can be found in \cite{Gal2016Uncertainty}.


\bibliographystyle{mnras}
\bibliography{bibliography}


\bsp	
\label{lastpage}
\end{document}